\documentclass[twocolumn,showpacs,preprintnumbers,amsmath,amssymb,prd,footinbib,floatfix]{revtex4-2}

\usepackage{graphicx}

\usepackage[pdftex, pdftitle={Localization of Dirac modes in the
  ${\rm SU}(2)$ Higgs model at finite temperature}]{hyperref}

\newcommand{\f}[2]{\frac{#1}{#2}}
\newcommand{\tf}[2]{{\textstyle\f{#1}{#2}}}

\newcommand{\la}{\langle}

\newcommand{\ra}{\rangle}

\newcommand{\tr}{{\rm tr}\,}

\newcommand{\Vc}{{\cal V}}

\newcommand{\T}{{\rm T}}

\begin{document}

\title{Localization of Dirac modes in the $\mathrm{SU}(2)$ Higgs model at finite temperature}

\author{Gy{\"o}rgy Baranka}
\email{barankagy@caesar.elte.hu}
\affiliation{ELTE E\"otv\"os Lor\'and University, Institute for
  Theoretical Physics, P\'azm\'any P\'eter s\'et\'any 1/A, H-1117, Budapest,
  Hungary}

\author{Matteo Giordano}
\email{giordano@bodri.elte.hu}
\affiliation{ELTE E\"otv\"os Lor\'and University, Institute for
  Theoretical Physics, P\'azm\'any P\'eter s\'et\'any 1/A, H-1117, Budapest,
  Hungary}

\begin{abstract}
  We investigate the connection between localization of low-lying
  Dirac modes and Polyakov-loop ordering in the lattice
  $\mathrm{SU}(2)$ Higgs model at finite temperature, probed with the
  staggered Dirac operator. After mapping out the phase diagram of
  the model at a fixed temporal extension in lattice units, we study
  the localization properties of the low-lying modes of the staggered
  Dirac operator, how these properties change across the various
  transitions, and how these modes correlate with the gauge and Higgs
  fields. We find localized low modes in the deconfined and in the
  Higgs phase, where the Polyakov loop is strongly ordered, but in
  both cases they disappear as one crosses over to the confined
  phase. Our findings confirm the general expectations of the
  ``sea/islands'' picture, and the more detailed expectations of its
  refined version concerning the favorable locations of localized
  modes, also in the presence of dynamical scalar matter.
\end{abstract}

\maketitle

\section{Introduction}
\label{sec:intro}

Although it is well established that the finite-temperature QCD
transition is an analytic
cross\-over~\cite{Aoki:2006we,Bazavov:2011nk}, the microscopic
mechanism that drives it is still being actively studied.  The main
goals of this line of research are a better understanding of the
connection between deconfinement and restoration of chiral symmetry,
both taking place in the crossover region; and of the fate of the
anomalous $\mathrm{U}(1)_A$ symmetry, especially in the chiral limit.
In this context, the fact that also the nature of the low-lying Dirac
eigenmodes changes radically in the crossover region has aroused some
interest. While delocalized in the low-temperature, confined and
chirally broken phase, these modes become in fact spatially localized
in the high-temperature, deconfined and (approximately) chirally
restored phase, up to a critical point in the spectrum known as
``mobility
edge''~\cite{GarciaGarcia:2006gr,Kovacs:2012zq,Giordano:2013taa,
  Ujfalusi:2015nha,Cossu:2016scb,Holicki:2018sms,Kehr:2023wrs} (see
Ref.~\cite{Giordano:2021qav} for a recent review). As the strength of
chiral symmetry breaking is controlled by the density of low-lying
Dirac modes~\cite{Banks:1979yr}, while the change in their
localization properties is mainly due to the ordering of the Polyakov
loop in the high-temperature phase~\cite{Bruckmann:2011cc,
  Giordano:2015vla,Giordano:2016cjs,Giordano:2016vhx,Giordano:2021qav,
  Baranka:2022dib}, low-lying eigenmodes could provide the link
between deconfinement and restoration of chiral symmetry.

The connection between low-mode localization and Polyakov-loop
ordering is qualitatively explained by the ``sea/islands'' picture,
initially proposed in Ref.~\cite{Bruckmann:2011cc}, and further
developed in Refs.~\cite{Giordano:2015vla,Giordano:2016cjs,
  Giordano:2016vhx,Giordano:2021qav,Baranka:2022dib}. In the
deconfined phase, typical gauge configurations display a ``sea'' of
ordered Polyakov loops, which on the one hand provides a spatially
(approximately) uniform region where Dirac modes can easily
delocalize, and on the other hand opens a (pseudo)gap in the near-zero
spectrum.  Polyakov-loop fluctuations away from order, and more
generally gauge-field fluctuations with reduced correlation in the
temporal direction, allow for eigenvalues below the gap; since in the
deconfined phase these fluctuations typically form well separated
``islands'', they tend to ``trap'' the low eigenmodes, causing their
localization.

The sea/islands mechanism is quite general, and requires essentially
only the ordering of the Polyakov loop for low-mode localization to
take place~\footnote{A notable exception is when the (untraced)
  Polyakov loops order along minus the identity, in which case low
  modes remain delocalized also in the deconfined
  phase~\cite{Baranka:2021san}.}.  This leads one to expect
localization of low Dirac modes to be a generic phenomenon in the
deconfined phase of a gauge theory, an expectation so far fully
confirmed by numerical results, both for pure gauge
theories~\cite{Gockeler:2001hr,Gattringer:2001ia,Gavai:2008xe,
  Kovacs:2009zj,Kovacs:2010wx,Bruckmann:2011cc, Kovacs:2017uiz,
  Giordano:2019pvc,Vig:2020pgq,Bonati:2020lal,Baranka:2021san,
  Baranka:2022dib} and in the presence of dynamical fermionic
matter~\cite{Giordano:2016nuu,Cardinali:2021fpu}. An interesting
aspect of the deconfinement/localization relation is that while the
thermal transition can be a smooth, analytic crossover, the appearance
of a mobility edge can only be sudden, taking place at a well-defined
temperature. If the connection between deconfinement and localization
is indeed general, one can then associate the (possibly smooth)
thermal transition with a (definitely sharp) ``geometric'' transition
(a similar suggestion, although in connection with deconfinement and
center vortices, was made in Ref.~\cite{Ghanbarpour:2022oxt}, from
which we borrowed the terminology). This point of view is supported by
the fact that the geometric and the thermodynamic transition coincide
when the latter is a genuine phase transition~\cite{Kovacs:2017uiz,
  Giordano:2019pvc,Vig:2020pgq,Bonati:2020lal,Baranka:2021san,
  Baranka:2022dib,Giordano:2016nuu,Cardinali:2021fpu}.

As a further test of the universality of the sea/islands mechanism,
one can investigate whether a change in the localization properties of
low modes takes place across other thermal transitions where the
Polyakov loop gets ordered, besides the usual deconfinement
transition. As an example, Ref.~\cite{Bonati:2020lal} studied low-mode
localization across the ``reconfinement'' transition in trace-deformed
$\mathrm{SU}(3)$ gauge theory at finite
temperature~\cite{Myers:2007vc,
  Unsal:2008ch,Bonati:2018rfg,Bonati:2019kmf,Athenodorou:2020clr}. While
localized modes are present in the deconfined phase also at nonzero
deformation parameter, where the Polyakov-loop expectation value is
different from zero, they disappear as the system reconfines and the
Polyakov-loop expectation value vanishes.

Yet another test of universality consists in changing the type of
dynamical matter from fermionic to scalar. As long as a phase with
ordered Polyakov loops exists, this should not affect the expectations
of the sea/islands picture, and localized modes should appear in the
spectrum of the Dirac operator in that phase. In this context, the
Dirac operator can be seen simply as a mathematical probe of certain
properties of the gauge fields or, more physically, as a probe of how
these fields couple to external, static (i.e., infinitely heavy)
fermion fields.

A model allowing one to carry out both these tests at once is the
lattice fixed-length $\mathrm{SU}(2)$ Higgs
model~\cite{Fradkin:1978dv}. At zero temperature the phase diagram of
this model has been studied in depth both with
analytical~\cite{Fradkin:1978dv} and
numerical~\cite{Lang:1981qg,Montvay:1984wy,Montvay:1985nk,Langguth:1985eu,
  Campos:1997dc,Bonati:2009pf} methods. This model has two parameters,
namely the (inverse) gauge coupling $\beta$ and the Higgs-gauge
coupling $\kappa$, and it displays two lines of transitions in the
$(\beta,\kappa)$ plane as follows~\cite{Bonati:2009pf}:
\begin{itemize}
\item a line of crossovers at $\beta\approx \beta_{\mathrm{bulk}}$,
  starting from the bulk transition (crossover) of the pure gauge
  $\mathrm{SU}(2)$ theory~\cite{Lautrup:1980dg} at
  $(\beta,\kappa)=(\beta_{\mathrm{bulk}},0)$, and ending at some point
  $(\beta_e,\kappa_e)$;
\item a line of crossovers coming down from large $\kappa$ at small
  $\beta$, meeting the first line at $(\beta_e,\kappa_e)$, turning
  into a line of first-order transitions at $(\beta_f,\kappa_f)$, and
  tending to $\kappa\approx 0.6$ as $\beta\to \infty$.
\end{itemize}
These transition lines separate three phases of the system: a confined
phase at low $\beta$ and low $\kappa$; a deconfined phase at high
$\beta$ and low $\kappa$; and a Higgs phase at high $\kappa$. A
similar phase diagram was found at finite temperature, although the
transition lines were all identified as crossovers in that
case~\cite{Bonati:2009pf}.  The absence of a sharp transition between
the confined and the Higgs phase at any $\kappa$ at sufficiently low
$\beta$ was proved in Ref.~\cite{Fradkin:1978dv}, where it was also
shown that in this region all local correlation functions, and so the
spectrum of the theory, depend analytically on the couplings.

While fermions are absent in the $\mathrm{SU}(2)$ Higgs model, one can
still probe this system using static external fermions coupled to the
$\mathrm{SU}(2)$ gauge field, as pointed out above. One can then study
how the corresponding Dirac spectrum behaves, and check what happens
to the localization properties of its low modes across the various
transitions, in particular as one crosses over to the Higgs phase
starting from either the confined or the deconfined phase. Since
eigenvalues and eigenvectors of the Dirac operator are nonlocal
functions of the gauge fields, they can display non-analytic behavior
even in the strip of the $(\beta,\kappa)$ plane where all local
correlators are analytic functions of the couplings, and so they could
allow one to sharply distinguish the confined and the Higgs phase. (A
different approach to this issue, based on the analogies between
gauge-Higgs theories and spin glasses, is discussed in the review
Ref.~\cite{Greensite:2021fyi} and references therein.)

In this paper we study the spectrum and the eigenvectors of the
staggered lattice Dirac operator in the $\mathrm{SU}(2)$ Higgs model
at finite temperature. After briefly describing the model, in section
\ref{sec:su2h} we introduce the tools we use to investigate the
localization properties of staggered eigenmodes.  In section
\ref{sec:ftPD} we map out the phase diagram of the model at finite
temperature, working at fixed temporal extension in lattice units. In
section \ref{sec:loc} we analyze the staggered eigenmodes, focussing
in particular on how their localization properties change across the
transitions between the confined, deconfined, and Higgs phases. We
then study in detail the correlation between eigenmodes and the gauge
and Higgs fields, to identify the field fluctuations mostly
responsible for localization. Finally, in section \ref{sec:concl} we
draw our conclusions and show some prospects for the future.

\section{$\mathrm{SU}(2)$ Higgs model and localization}
\label{sec:su2h}

In this section we describe the fixed-length $\mathrm{SU}(2)$ Higgs
model, and discuss how to characterize the localization properties of
Dirac modes, and how these correlate with the gauge and Higgs fields.

\subsection{$\mathrm{SU}(2)$ Higgs model on the lattice}
\label{sec:su2hmodel}

We study the lattice $\mathrm{SU}(2)$ Higgs model in 3+1 dimensions,
defined by the action
\begin{equation}
  \label{eq:su2h1}
  S = - \f{\beta }{2}\sum_n\sum_{1\le \mu<\nu \le 4}\tr U_{\mu\nu}(n) -
  \f{\kappa}{2}\sum_n\sum_{1\le \mu\le 4} \tr G_\mu(n)
  \,,
\end{equation}
where we omitted an irrelevant additive constant. Here
$n=(\vec{x},t)$, $n_\mu=0,\ldots,N_\mu-1$, are the sites of a
hypercubic $N_s^3\times N_t$ lattice, i.e., $N_{1,2,3}=N_s$ and
$N_4=N_t$, where $\mu=1,\ldots, 4$ denotes the lattice directions and
$\hat{\mu}$ the corresponding unit vectors. The dynamical variables
are the $\mathrm{SU}(2)$ matrices $U_\mu(n)$ and $\phi(n)$,
representing respectively the gauge variables associated with the link
connecting $n$ and $n+\hat{\mu}$, and the unit-length Higgs field
doublet (recast as a unitary matrix) associated with site $n$, and
\begin{equation}
  \label{eq:su2h2}
  \begin{aligned}
  U_{\mu\nu}(n) &= U_\mu(n) U_\nu(n+\hat{\mu}) U_\mu(n+\hat{\nu})^\dag
  U_\nu(n)^\dag\,,\\
  G_\mu(n) &=  \phi(n)^\dag U_\mu(n) \phi(n+\hat{\mu})\,,
  \end{aligned}
\end{equation}
are the plaquette variables associated with the elementary lattice
squares, and the nontrivial part of the discretized covariant
derivative of the Higgs field, which we will refer to as the
Higgs-gauge field coupling term. Periodic boundary conditions are
imposed on $U_\mu(n)$ and $\phi(n)$ in all directions. In what
follows we will also make use of the Polyakov loop winding around the
temporal direction,
\begin{equation}
  \label{eq:su2h2_bis}
  P(\vec{x}) = \tr \prod_{t=0}^{N_t-1}
  U_4(\vec{x},t)\,.
\end{equation}
Expectation values are defined as
\begin{equation}
  \label{eq:su2h_expval}
  \begin{aligned}
  \la O \ra &= \f{1}{Z}\int DU \int D\phi
  \,e^{-S(U,\phi)}O(U,\phi)\,, \\
  Z &= \int DU \int D\phi
  \,e^{-S(U,\phi)}\,,
  \end{aligned}
\end{equation}
where $DU$ and $D\phi$ denote the products of the $\mathrm{SU}(2)$
Haar measures associated with $U_\mu(n)$ and $\phi(n)$.

We study this model at finite temperature $T=1/(aN_t)$, where $a$ is
the lattice spacing, which can be set by suitably tuning the
parameters of the model, namely the inverse gauge coupling $\beta$ and
the Higgs-gauge field coupling $\kappa$.  However, since we are not
interested here in taking the continuum limit, we treat the model
simply as a two-parameter anisotropic statistical mechanics system,
keeping $N_t$ fixed as we take the thermodynamic limit
$N_s\to \infty$, and as we change $\beta$ and $\kappa$ freely.
To study the phase diagram in the $(\beta,\kappa)$ plane we use the
average plaquette, Polyakov loop, and Higgs-gauge field coupling term,
\begin{equation}
  \label{eq:su2h7}
  \begin{aligned}
    \la U\ra &= \f{1}{N_t V}\sum_{n}\left\la U(n) \right\ra\,,     &&&
    \la P\ra &= \f{1}{V}\sum_{\vec{x}}\left\la P(\vec{x}) \right\ra\,,
    \\
    \la G\ra &= \f{1}{N_t V}\sum_{n}\left\la G(n) \right\ra\,,
  \end{aligned}
\end{equation}
where $V=N_s^3$ is the lattice volume, and the corresponding
susceptibilities,
\begin{equation}
  \label{eq:su2h8}
  \begin{aligned}
    \chi_U &= \f{1}{N_t V}\left(\left\la \left({\textstyle\sum_{n}}
          U(n)\right)^2\right\ra -\left\la
        {\textstyle\sum_{n}} U(n)\rule{0pt}{12pt}\right\ra^2\right)\,,\\
    \chi_P &= \f{1}{ V}\left(\left\la
        \left({\textstyle\sum_{\vec{x}}}\, \tr
          P(\vec{x})\right)^2\right\ra -\left\la
        {\textstyle\sum_{\vec{x}}}\,\tr P(\vec{x})\rule{0pt}{12pt}\right\ra^2\right)\,,\\
    \chi_G &= \f{1}{N_t V}\left(\left\la \left({\textstyle\sum_{n}}
          G(n) \right)^2\right\ra -\left\la {\textstyle\sum_{n}}\tr
        G(n)\rule{0pt}{12pt}\right\ra^2\right)\,.
  \end{aligned}
\end{equation}
In Eqs.~\eqref{eq:su2h7} and \eqref{eq:su2h8} we denoted with $U(n)$
and $G(n)$ the average plaquette and gauge-Higgs coupling term
touching a lattice site $n$,
\begin{equation}
  \label{eq:su2h7_alt2}
  \begin{aligned}
    U(n) &= \f{1}{24}\sum_{1\le\mu<\nu\le 4} \tr (U_{\mu\nu}(n)+
    U_{\mu\nu}(n-\hat{\mu})
 \\ & \phantom{\f{1}{24}\sum_{1\le\mu<\nu\le 4} }
    + U_{\mu\nu}(n-\hat{\nu})  +
    U_{\mu\nu}(n-\hat{\mu}-\hat{\nu}) ) \,,
    \\
    G(n) &= \f{1}{8}\sum_{1\le\mu\le 4} \tr (G_\mu(n)+
    G_\mu(n-\hat{\mu})) \,.
  \end{aligned}
\end{equation}

\subsection{Localization of staggered eigenmodes}
\label{sec:stloc}

We are interested in the spectrum of the staggered Dirac operator in
the background of the $\mathrm{SU}(2)$ gauge fields for fermions in
the fundamental representation,
\begin{equation}
  \label{eq:su2h3}
  D^{\mathrm{stag}} = \f{1}{2}\sum_{\mu} \eta_\mu (U_\mu \T_\mu -
  \T_\mu^\dag U_\mu^\dag)\,, 
\end{equation}
where $\eta_\mu$ are the usual staggered phases and $\T_\mu$ are the
translation operators with periodic (resp.\ antiperiodic) boundary
conditions in space (resp.\ time), i.e.,
\begin{equation}
  \label{eq:su2h4}
  \begin{aligned}
  \eta_{\mu}(n) &= (-1)^{\sum_{\alpha <\mu} n_\alpha}\,, \\
  (\T_\mu)_{n,n'} &= b_\mu(n_\mu)\delta_{n_\mu+1,
    n_\mu'}\prod_{\alpha\ne \mu} \delta_{n_\alpha, n_\alpha'}\,, 
  \end{aligned}
\end{equation}
with $n_\mu=N_\mu$ identified with $n_\mu=0$, and $b_\mu(n_\mu)=1$,
$\forall\mu,n_\mu$, except for $b_4(N_t-1)=-1$. Since the staggered
operator is anti-Hermitian and anticommutes with
$\varepsilon(n)=(-1)^{\sum_{\alpha} n_\alpha}$, its spectrum is purely
imaginary and symmetric about the origin. We write
\begin{equation}
  \label{eq:su2h5}
  D^{\mathrm{stag}}\psi_l(n) = i\lambda_l \psi_l(n)\,, \qquad
  \lambda_l \in\mathbb{R}\,,
\end{equation}
with eigenvectors $\psi_l(n)$ carrying an internal ``color'' index,
$\psi_{l,c}(n)$, $c=1,2$, that has been suppressed for simplicity, and
focus on $\lambda_l\ge 0$ only. Notice that since
$\sigma_2 U_\mu(n) \sigma_2=U_\mu(n)^*$, $D^{\mathrm{stag}}$ commutes
with the antiunitary ``time-reversal'' operator $ T = \sigma_2 K$,
where $K$ denotes complex conjugation. Since $ T^2= -\mathbf{1}$,
$D^{\mathrm{stag}}$ displays in this case doubly degenerate
eigenvalues, and belongs to the symplectic class in the symmetry
classification of random
matrices~\cite{mehta2004random,Verbaarschot:2000dy}.  In the following
it is understood that we work with the reduced spectrum, including
only one eigenvalue from each degenerate pair.

\paragraph{Participation ratio}

The localization properties of the staggered eigenmodes can be studied
directly by looking at the eigenvectors, or indirectly by looking at
the corresponding eigenvalues. In the first case one can study the
volume scaling of the so-called participation ratio (PR) of the modes,
\begin{equation}
  \label{eq:su2h9}
  \mathrm{PR}_l =  
  \f{1}{N_t V}  \mathrm{IPR}_l^{-1}\,, \qquad
  \mathrm{IPR}_l = \sum_n \Vert \psi_l(n)\Vert^4\,, 
\end{equation}
where $\Vert \psi_l(n)\Vert^2=\sum_{c=1}^2|\psi_{l,c}(n)|^2$, modes
are normalized to 1, $\sum_n \Vert \psi_l(n)\Vert^2=1$, and IPR is the
inverse participation ratio. The quantity $\mathrm{PR}_l$ measures the
fraction of lattice volume $N_t V$ occupied by a given mode, and
similarly $N_t V\cdot \mathrm{PR}_l=\mathrm{IPR}_l^{-1}$ gives the
``mode size''.  After averaging over an infinitesimally small spectral
bin around a point $\lambda$ in the spectrum and over gauge
configurations, as the spatial size $N_s$ grows the resulting average
$\mathrm{PR}(\lambda,N_s)$ tends to a constant if modes near $\lambda$
are delocalized on the entire lattice, and goes to zero as the inverse
of the lattice volume if they are localized in a finite region.
Equivalently, the similarly averaged mode size diverges linearly in
the lattice volume for delocalized modes and tends to a constant for
localized modes. In this paper we denote the average of any
observable $O_l$ associated with mode $l$, following the procedure
described above, as
\begin{equation}
  \label{eq:su2h10}
  O(\lambda,N_s) =
  \f{\left\la\sum_l\delta(\lambda-\lambda_l) O_l
    \right\ra}{\left\la\sum_l\delta(\lambda-\lambda_l)\right\ra}\,, 
\end{equation}
having made explicit the dependence on the spatial size of the
lattice.  The volume scaling of $\mathrm{PR}(\lambda,N_s)$ defines the
fractal dimension of modes in the neighborhood of $\lambda$,
\begin{equation}
  \label{eq:fracdim}
  \alpha(\lambda) = 3+\lim_{N_s\to\infty} \f{\log
    \mathrm{PR}(\lambda,N_s)}{\log N_s}\,. 
\end{equation}
The multifractal properties of eigenmodes can be investigated by
looking at the generalized inverse participation ratios,
\begin{equation}
  \label{eq:genIPR}
 (\mathrm{IPR}_q)_l = \sum_n \Vert \psi_l(n)\Vert^{2q}\,,
\end{equation}
with $(\mathrm{IPR}_2)_l=\mathrm{IPR}_l$~\footnote{Since
   \protect{$\Vert T\psi_l(n)\Vert^2=\Vert \psi_l(n)\Vert^2$},
  eigenmodes in a
  degenerate pair have identical $(\mathrm{IPR}_q)_l$.}. Their average
according to Eq.~\eqref{eq:su2h10} scales with the system size as
$\mathrm{IPR}_q(\lambda,N_s) \propto N_s^{-D_q(\lambda)(q-1)}$, with
generalized fractal dimensions $D_q$ (notice $D_2 = \alpha$). One has
$D_q=3$ for delocalized modes and $D_q=0$ for localized modes, while a
nontrivial $D_q$ signals eigenmode
multifractality~\cite{Evers:2008zz}.

\paragraph{Spectral statistics}

The localization properties of the eigenmodes reflect on the
statistical properties of the
eigenvalues~\cite{altshuler1986repulsion}: for localized modes one
expects independent fluctuations of the eigenvalues, while for
delocalized modes one expects to find the correlations typical of
dense random matrix models. It is convenient in this context to study
the probability distribution of the so-called unfolded level
spacings~\cite{mehta2004random,Verbaarschot:2000dy},
\begin{equation}
  \label{eq:su2h11}
  s_l = \f{\lambda_{l+1}-\lambda_l}{\la\lambda_{l+1}-\lambda_l\ra_\lambda }\,,
\end{equation}
computed locally in the spectrum, i.e.,
\begin{equation}
  \label{eq:su2h12}
  p(s;\lambda,N_s) =
  \f{\left\la\sum_l\delta(\lambda-\lambda_l)\delta(s-s_l)\right\ra}{\left\la\sum_l
      \delta(\lambda-\lambda_l)\right\ra}\,. 
\end{equation}
In Eq.~\eqref{eq:su2h11}, $\la\lambda_{l+1}-\lambda_l\ra_\lambda$
denotes the average spacing in the relevant spectral region, which for
large volumes equals
$\la\lambda_{l+1}-\lambda_l\ra_\lambda \to \f{1}{N_t V\rho(\lambda)}$,
where $\rho(\lambda)$ is the spectral density,
\begin{equation}
  \label{eq:su2h13}
  \rho(\lambda) =  \lim_{V\to \infty}
  \f{1}{N_t V}
  \left\la{\textstyle\sum_l}\delta(\lambda-\lambda_l)
  \right\ra\,.
\end{equation}
The statistical properties of the unfolded spacings are expected to be
universal~\cite{mehta2004random}, i.e., independent of the details of
the model, and can be compared to the theoretical predictions obtained
from exactly solvable models.  As the system size increases, for
localized modes $p(s;\lambda,N_s)$ should approach the exponential
distribution, $p_{\mathrm{P}}(s)=e^{-s}$, appropriate for independent
eigenvalues obeying Poisson statistics~\cite{mehta2004random}. For
delocalized modes $p(s;\lambda,N_s)$ should instead approach the
distribution $p_{\mathrm{RMT}}(s)$ predicted by the appropriate
Gaussian Ensemble of Random Matrix Theory, which is the Gaussian
Symplectic Ensemble in the case at
hand~\cite{mehta2004random,Verbaarschot:2000dy}. This quantity is
known exactly, but is not available in closed form. An accurate
approximation is provided by the symplectic Wigner surmise,
\begin{equation}
  \label{eq:symWS}
  p_{\mathrm{WS}}(s) =\left(\f{64}{9\pi}\right)^3s^4e^{-\f{64}{9\pi} s^2}\,.
\end{equation}

\paragraph{Mobility edge}

Localized and delocalized modes are generally found in disjoint
spectral regions separated by critical points known as
\textit{mobility edges}, where the localization length diverges and
the system undergoes a phase transition along the spectrum, known as
Anderson transition~\cite{Evers:2008zz}. At the mobility edge the
critical eigenmodes display a fractal dimension different from those
of localized or delocalized modes, as well as a rich multifractal
structure. This is reflected in critical spectral statistics different
from both Poisson and RMT statistics.  To monitor how the localization
properties change along the spectrum using its statistical properties,
it is convenient to use the integrated unfolded level spacing
distribution,
\begin{equation}
  \label{eq:su2h14}
  I_{s_0}(\lambda,N_s) = \int_0^{s_0} ds\,
  p(s;\lambda,N_s)\,,
\end{equation}
where $s_0\simeq 0.563$ is chosen so to maximize the difference
between the expectations for Poisson and RMT distributions,
$I_{s_0,\mathrm{P}}\simeq 0.431$ and
$I_{s_0,\mathrm{RMT}}\simeq 0.0797$, estimated using $p_{\mathrm{P}}$
and $p_{\mathrm{WS}}$, see Eq.~\eqref{eq:symWS} above.  This quantity
allows one to determine the mobility edge very accurately by means of
a finite-size-scaling analysis~\cite{Shklovskii:1993zz}.  In fact, as
the system size increases $I_{s_0}(\lambda,N_s)$ tends to
$I_{s_0,\mathrm{P}}$ or $I_{s_0,\mathrm{RMT}}$ depending on the
localization properties of the modes in the given spectral region,
except at the mobility edge where it is volume-independent and takes
the value $I_{s_0,c}$ corresponding to the critical statistics.  This,
however, requires large-scale simulations to achieve a sufficient
quality of the data, and several large volumes.

One can give up some of the accuracy but save a lot in computing
effort by using the critical value of the spectral statistic, expected
to be universal, to determine the mobility edge simply by looking for
the point where the curve for $I_{s_0}$ crosses its critical value,
$I_{s_0,c}$ (see, e.g.,
Refs.~\cite{Giordano:2016nuu,Kovacs:2017uiz,Vig:2020pgq,
  Cardinali:2021fpu}). This critical value is not known for the
symplectic class, but it can be determined by identifying the
scale-invariant point in the spectrum at some point in the parameter
space of the model under study (if one can find an Anderson
transition, of course); the corresponding critical value can then be
used in the rest of the analysis. Notice that one could estimate the
mobility edge in a finite volume as the point where $I_{s_0}$ takes
any chosen value intermediate between the RMT and the Poisson
predictions, and this would converge to the correct value in the
infinite-volume limit. In this respect, the choice of $I_{s_0,c}$ is
only the most convenient, as it is expected to minimize the magnitude
of finite-size effects.

\paragraph{Correlation with bosonic observables}

To investigate the correlation between staggered eigenmodes and gauge
and Higgs fields we considered the following observables,
\begin{equation}
  \label{eq:su2h15}
  \begin{aligned}
    \mathcal{U}_l &=\sum_{n} U(n)
    \Vert\psi_l(n)\Vert^2\,,&&& \mathcal{P}_l &= \sum_{t,\vec{x}}
    P(\vec{x}) \Vert\psi_l(\vec{x},t)\Vert^2\,,\\ \mathcal{G}_l &=
    \sum_{n} G(n) \Vert\psi_l(n)\Vert^2\,,
  \end{aligned}
\end{equation}
averaged according to Eq.~\eqref{eq:su2h10}. Recall that $U(n)$ and
$G(n)$ are the average plaquette and gauge-Higgs coupling term
touching a lattice site $n$, defined in Eq.~\eqref{eq:su2h7_alt2}. For
delocalized modes $\Vert\psi_l\Vert^2\sim \f{1}{V}$, and the averages
$\mathcal{U}(\lambda,N_s)$, $\mathcal{P}(\lambda,N_s)$, and
$\mathcal{G}(\lambda,N_s)$ of the observables in Eq.~\eqref{eq:su2h15}
are approximately equal to the average of the corresponding bosonic
observable, i.e., $\la U\ra$, $\la P\ra$, and $\la G\ra$, respectively
[see Eq.~\eqref{eq:su2h7}]. For localized modes $\Vert\psi_l\Vert^2$
is non-negligible only inside a region of finite spatial volume, so
$\mathcal{P}(\lambda,N_s)$ measures the average Polyakov loop inside
the localization region, and $\mathcal{U}(\lambda,N_s)$ and
$\mathcal{G}(\lambda,N_s)$ measure respectively the average plaquette
and gauge-Higgs coupling term in a neighborhood of the localization
region. One should, however, keep in mind that there are 24
neighboring squares and 8 neighboring links to each site, so that a
possible correlation of modes with the plaquette and gauge-Higgs
coupling term fluctuations get diluted.

More informative than the averages of the observables in
Eq.~\eqref{eq:su2h15} are the corresponding centered and rescaled
averages,
\begin{equation}
  \label{eq:su2h15bis}
  \begin{aligned}
    \widehat{\mathcal{U}}(\lambda,N_s) &= \f{
      \mathcal{U}(\lambda,N_s)-\la U\ra }{\delta U}\,, \\
    (\delta U)^2 &=   \la U(n)^2 \ra - \la U(n) \ra^2\,,\\
    \widehat{\mathcal{P}}(\lambda,N_s) &= \f{
      \mathcal{P}(\lambda,N_s)-\la P\ra }{\delta P}\,, \\
    (\delta P)^2 &=   \la P(\vec{x})^2 \ra - \la P(\vec{x}) \ra^2\,,\\
    \widehat{\mathcal{G}}(\lambda,N_s) &= \f{
      \mathcal{G}(\lambda,N_s)-\la G\ra }{\delta G}\,, \\ (\delta
    G)^2 &= \la G(n)^2 \ra - \la G(n) \ra^2\,.
  \end{aligned}
\end{equation}
These quantities measure the correlation of the eigenmodes with
fluctuations in the gauge and Higgs fields, normalized by the average
size of these fluctuations. Indeed, writing these quantities out
explicitly, one has, e.g.,
\begin{equation}
  \label{eq:su2h15ter}
  \begin{aligned}
    \widehat{\mathcal{U}}(\lambda,N_s)= \left\la{\textstyle\sum_n}
      \f{{\textstyle\sum_l} \delta(\lambda-\lambda_l)\Vert
        \psi_l(n)\Vert^2}{N_t V\rho(\lambda)} \,\f{U(n)- \la
        U\ra}{\delta U} \right\ra\,.
  \end{aligned}
\end{equation}
As a consequence, the observables in Eq.~\eqref{eq:su2h15bis} vanish
in the absence of correlation, and are strongly suppressed for
delocalized modes.  The normalization factor takes into account that
for observables with a strongly peaked probability distribution even a
correlation with small deviations from average is significant,
indicating that eigenmodes are attracted by the corresponding type of
fluctuations, and favor the locations where they show up in a field
configuration. In particular, for localized modes this allows one to
identify the most favorable type of fluctuations for localization.

\paragraph{Sea/islands picture}

We also study the correlation between eigenmodes and the ``islands''
of the refined ``sea/is\-lands'' picture of localization discussed in
Ref.~\cite{Baranka:2022dib}. These are defined using the
``Dirac-Anderson Hamiltonian'' representation of the staggered Dirac
operator~\cite{Giordano:2016cjs}, obtained by diagonalizing the
temporal hopping term in $D^{\mathrm{stag}}$ [i.e., the term with
$\mu=4$ in the sum in Eq.~\eqref{eq:su2h3}] by means of a unitary
transformation $\Omega$~\cite{Baranka:2022dib},
\begin{equation}
  \label{eq:si1}
  H^{\mathrm{DA}} \equiv \Omega^\dag (-iD^{\mathrm{stag}})\Omega =
  \mathcal{E}  \mathbf{1}_s     + \f{1}{2i}\sum_{j=1}^3 \eta_j
  (\Vc_j \T_j - \T_j^\dag \Vc_j^\dag)\,,
\end{equation}
where $\mathbf{1}_s$ is the $V\times V$ identity matrix
$(\mathbf{1}_s)_{\vec{x},\vec{y}}= \delta_{\vec{x},\vec{y}}$, $\T_j$
are here the spatial translation operators
$(\T_j)_{\vec{x},\vec{y}}=\delta_{\vec{x}+\hat{\jmath},\vec{y}}$ (with
periodic boundary conditions understood),
$\mathcal{E}$ is an $\vec{x}$-dependent $2N_t\times 2N_t$
diagonal matrix, 
\begin{equation}
  \label{eq:si2}
  \begin{aligned}
    \mathcal{E}(\vec{x})_{ka\,lb} &=
    \delta_{kl}\delta_{ab}e_{ka}(\vec{x})\,, &&& e_{ka}(\vec{x})&=
    \eta_4(\vec{x})\sin\omega_{ka}(\vec{x})\,,
  \end{aligned}
\end{equation}
and $\Vc_j$ are $\vec{x}$-dependent $2N_t\times 2N_t$ unitary
matrices,
\begin{equation}
  \label{eq:si2_bis}
  \begin{aligned}
    \Vc_j(\vec{x})_{ka\,lb} &=
    \f{1}{N_t}\sum_{t=0}^{N_t-1}e^{-i\left(\omega_{ka}(\vec{x}) -
        \omega_{lb}(\vec{x}+\hat{\jmath})\right)t}\, U_{j}^{\rm
      tdg}(\vec{x},t)_{ab}\,,
  \end{aligned}
\end{equation}
with $k,l=0,\ldots N_t-1$ and $a,b=1,2$. Here ``tdg'' stands for
``temporal diagonal gauge'', i.e., $U_{j}^{\rm tdg}$ are the spatial
links in the temporal gauge where all Polyakov loops are
diagonal~\footnote{In the definition of $U_{j}^{\rm tdg}$ in Eq.~(B.8)
  of Ref.~\cite{Baranka:2022dib} the factors $u(\vec{x})$ and
  $u(\vec{x}+\hat{\jmath})^\dag$ are incorrectly missing.},
\begin{equation}
  \label{eq:si3}
  \begin{aligned}
    U_{j}^{\rm tdg}(\vec{x},t) &=u(\vec{x})^\dag P(\vec{x},t)
    U_{j}(\vec{x},t)
    P(\vec{x}+ \hat{\jmath},t)^\dag u(\vec{x}+ \hat{\jmath})\,,\\
    P(\vec{x},t+1) &= P(\vec{x},t)U_4(\vec{x},t)\,,
  \end{aligned}
\end{equation}
with $P(\vec{x},0)=\mathbf{1}$, and $u(\vec{x})$ a suitable unitary
matrix such that [notice $P(\vec{x}) = P(\vec{x},N_t)$]
\begin{equation}
  \label{eq:si3_bis}
  P(\vec{x}) 
  = u(\vec{x})
  \mathrm{diag}(e^{i\phi_1(\vec{x})},e^{i\phi_2(\vec{x})})
  u(\vec{x})^\dag\,,
\end{equation}
with $\phi_{1,2}(\vec{x})\in[-\pi,\pi)$ and
$e^{i\left(\phi_1(\vec{x})+\phi_2(\vec{x})\right)}=1$. Moreover,
$\omega_{ka}(\vec{x})$ are effective Matsubara frequencies,
\begin{equation}
  \label{eq:si4}
  \omega_{ka}(\vec{x}) = \f{\phi_a(\vec{x})+(2n_{ka}+1)\pi}{N_t}\,,
\end{equation}
with $n_{ka}\in\{0,\ldots,N_t-1\}$ chosen for each $a$ so that the
``energies'' $e_{ka}$ satisfy
$0\le e_{1a}(\vec{x})\le e_{2a}(\vec{x})\le \ldots \le e_{\f{N_t}{2}-1
  \,a}$, and $e_{k+\f{N_t}{2}\,a}(\vec{x})=-e_{ka}(\vec{x})$, for
$k=0,\ldots \f{N_t}{2}-1$. Notice that thanks to the simple relation
between $\phi_1$ and $\phi_2$, one has $e_{k1}=e_{k2}$.  This double
degeneracy is a consequence of the temporal hopping term being
invariant under the time-reversal transformation $T$ (see section \ref
{sec:stloc}). With this choice for $e_{ka}$, $H^{\mathrm{DA}}$ has
the general structure
\begin{equation}
  \label{eq:si5}
  \begin{aligned}
    H^{\mathrm{DA}} &=
    \begin{pmatrix}
      E & \mathbf{0} \\ \mathbf{0} & -E
    \end{pmatrix}
\\ &\phantom{=}    + \f{1}{2i}\sum_{j=1}^3 \eta_j\left[
      \begin{pmatrix}
        A_j & B_j \\ B_j & A_j
      \end{pmatrix}\T_j
      -\T_j{}^\dag
      \begin{pmatrix}
        A_j{}^\dag & B_j{}^\dag \\ B_j{}^\dag & A_j{}^\dag
      \end{pmatrix}\right]\,,
  \end{aligned}
\end{equation}
where $E,A_j,B_j$ are $N_t\times N_t$ matrices.

It was argued in Ref.~\cite{Baranka:2022dib} that sites where the
diagonal blocks $A_j$ are larger are the most favorable for the
localization of low modes in a phase where the Polyakov loops are
ordered. In general, spatial regions with larger $A_j$, which
correspond to lower correlation among spatial links on different time
slices, are expected to be favored by low modes; in an ordered phase
such regions are localized, and so lead to low-mode localization. One
can check this by looking at the correlation between modes and the
quantity
\begin{equation}
  \label{eq:si6}
  A (\vec{x})
  = \f{1}{6N_t}
  \sum_{j=1}^3 \tr
  A_j(\vec{x})^\dag A_j(\vec{x}) + \tr  A_j(\vec{x}-\hat{\jmath})^\dag
  A_j(\vec{x}-\hat{\jmath})\,, 
\end{equation}
i.e., using the observable
\begin{equation}
  \label{eq:si7}
  \mathcal{A}_l = \sum_{\vec{x}} 
  A(\vec{x})
  \sum_{t=0}^{N_t-1}\Vert \psi_l(\vec{x},t)\Vert^2\,,
\end{equation}
averaged according to Eq.~\eqref{eq:su2h10} to get
$\mathcal{A}(\lambda,N_s)$, and centered and rescaled according to
Eq.~\eqref{eq:su2h15ter} to get $\widehat{\mathcal{A}}(\lambda,N_s)$,
i.e.,
\begin{equation}
  \label{eq:si8}
  \begin{aligned}
    \widehat{\mathcal{A}}(\lambda,N_s) &= \f{ \mathcal{A}(\lambda,N_s)-\la
      A\ra }{\delta A}\,, \\
    (\delta A)^2 &=   \la A(\vec{x})^2 \ra - \la A(\vec{x}) \ra^2\,.
  \end{aligned}
\end{equation}

\section{Phase diagram at finite temperature}
\label{sec:ftPD}

In this section we report our results on the phase diagram of the
model. We worked at finite temperature, fixing the lattice temporal
extension to $N_t=4$, and performing numerical simulations with a
standard heatbath algorithm.

Theoretical arguments~\cite{Fradkin:1978dv} and previous numerical
studies~\cite{Bonati:2009pf} lead us to expect three phases: a
confined phase at small $\beta$ and small $\kappa$; a deconfined phase
at large $\beta$ and small $\kappa$; and a Higgs phase at large
$\kappa$. Based on the finite-temperature results of
Ref.~\cite{Bonati:2009pf}, and on the observed weakening of the
transition for smaller temporal extensions reported there, we expect
that the transitions between the three phases are analytic
crossovers. A detailed study of this issue is beyond the scope of this
paper, so we limited most of our simulations to a single lattice
volume with $N_s=20$, for 784 different $(\beta,\kappa)$ pairs, using
3000 configurations at each point. We took $\kappa \in [0,1.35]$ in
steps of $\Delta\kappa=0.05$ and $\beta \in [1.5,2.85]$ in steps of
$\Delta\beta=0.05$. A detailed volume-scaling study was done on a
subset of these points: we discuss this below.

\begin{figure}[t]
  \centering
  \includegraphics[width=0.32\textwidth]{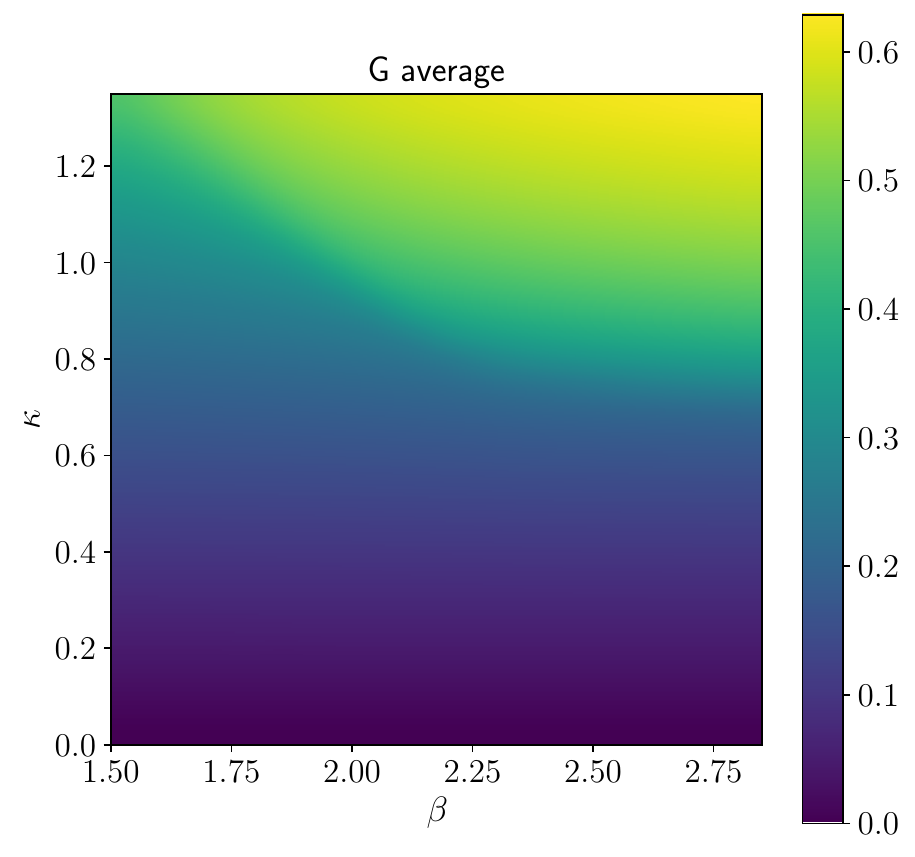}
  \hfil
  \includegraphics[width=0.33\textwidth]{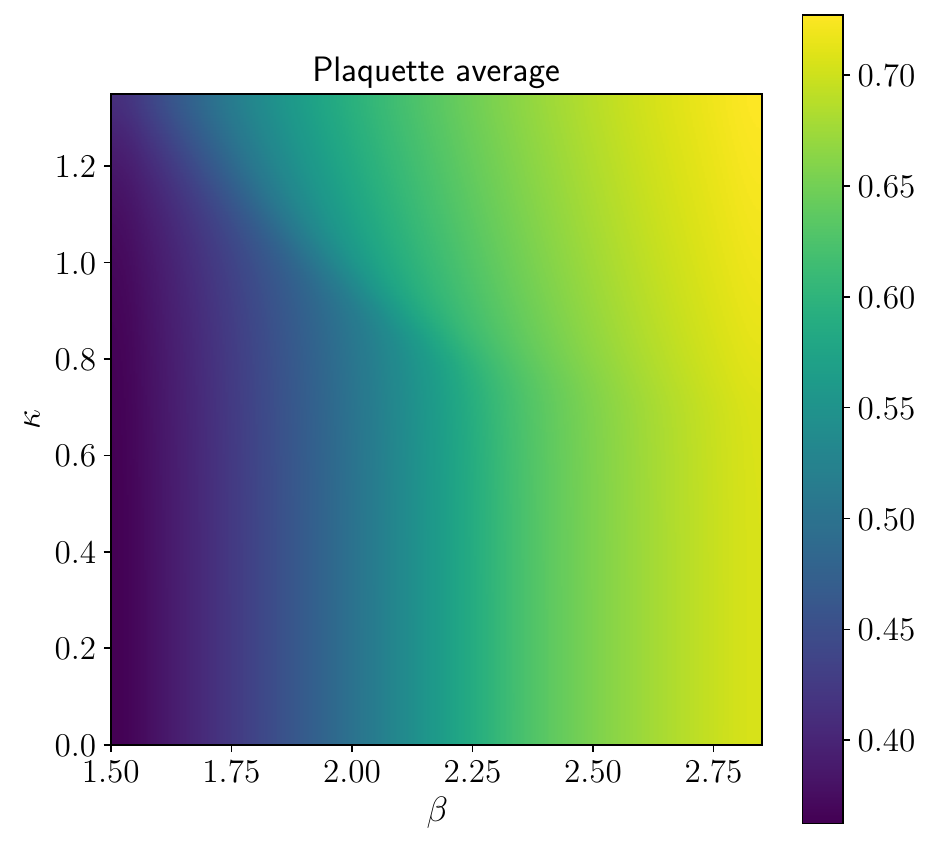}
  \hfil  
  \includegraphics[width=0.32\textwidth]{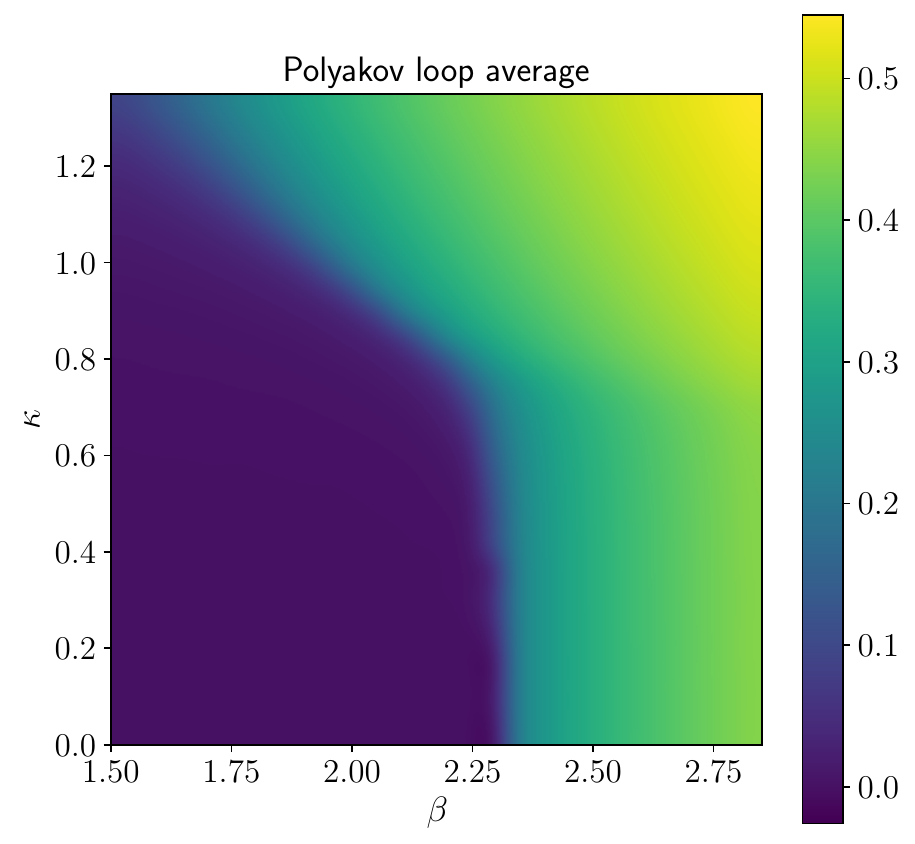}
  \caption{Heatmap plot of the expectation value of $G$ (top panel),
    $U$ (center panel), and $P$ (bottom panel), see
    Eq.~\eqref{eq:su2h7}. Here $N_s=20$ and $N_t=4$.}
  \label{fig:0}
\end{figure}

We show our results for $\la G\ra$, $\la U\ra$, and $\la P\ra$ in
Fig.~\ref{fig:0} as heatmap plots, obtained by cubic interpolation of
the numerical results at the simulation points. These confirm our
expectations, and allow us to characterize the confined phase at small
$\beta$ and $\kappa$ by small $\la G\ra$, $\la U\ra$, and $\la P\ra$;
the deconfined phase at large $\beta$ and small $\kappa$ by small
$\la G\ra$ and large $\la U\ra$ and $\la P\ra$; and the Higgs phase at
large $\kappa$ by large $\la G\ra$, $\la U\ra$, and $\la P\ra$. We
estimated errors with a standard jackknife procedure: they are not
shown, but relative errors are always within $7\cdot 10^{-5}$ for
$\la U\ra $; $2\cdot 10^{-3}$ for $\la G \ra$; and within
$1\cdot 10^{-3}$ for $\la P\ra$, except deep inside the confined phase
where the average becomes very small and indistinguishable from zero
within errors. More precisely, the expectation value of the
gauge-Higgs coupling term (Fig.~\ref{fig:0}, top panel) divides the
phase diagram into two pieces: the Higgs phase at large $\kappa$, with
large $\la G \ra$, and the (undivided) confined and deconfined phases
at small $\kappa$, with similar and small values of $\la G \ra$.  The
expectation value of the plaquette and of the Polyakov loop
(Fig.~\ref{fig:0}, center and bottom panel) divide the phase diagram
into two parts in a different way: the confined phase at low $\beta$
and $\kappa$, where both $\la U\ra$ and $\la P\ra$ are small, and the
(undivided) Higgs and deconfined phases, where both $\la U\ra$ and
$\la P\ra$ are large.

\begin{figure}[t]
  \centering
  \includegraphics[width=0.32\textwidth]{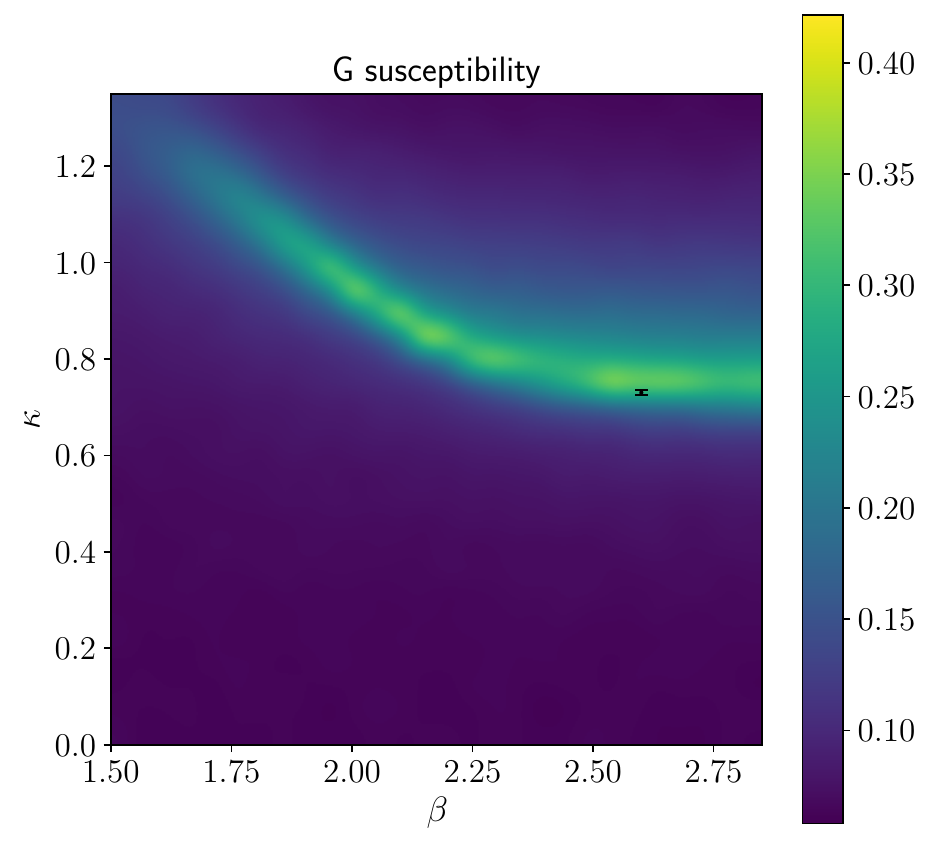}
  \hfil
  \includegraphics[width=0.32\textwidth]{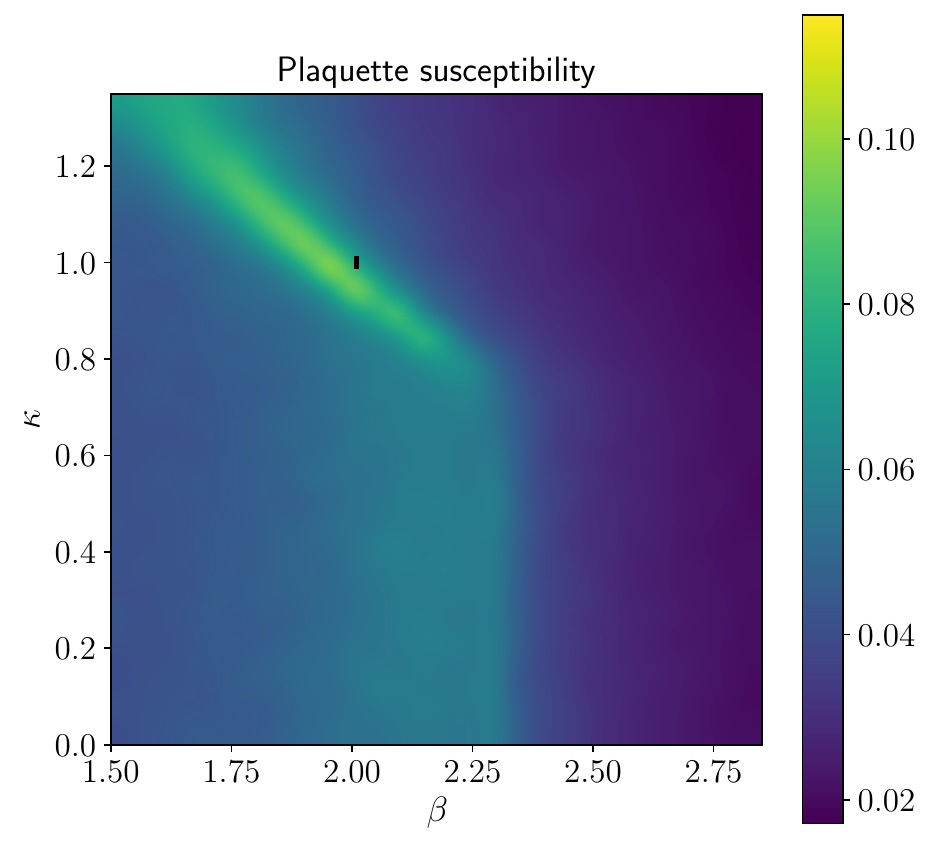}
  \hfil
    \includegraphics[width=0.32\textwidth]{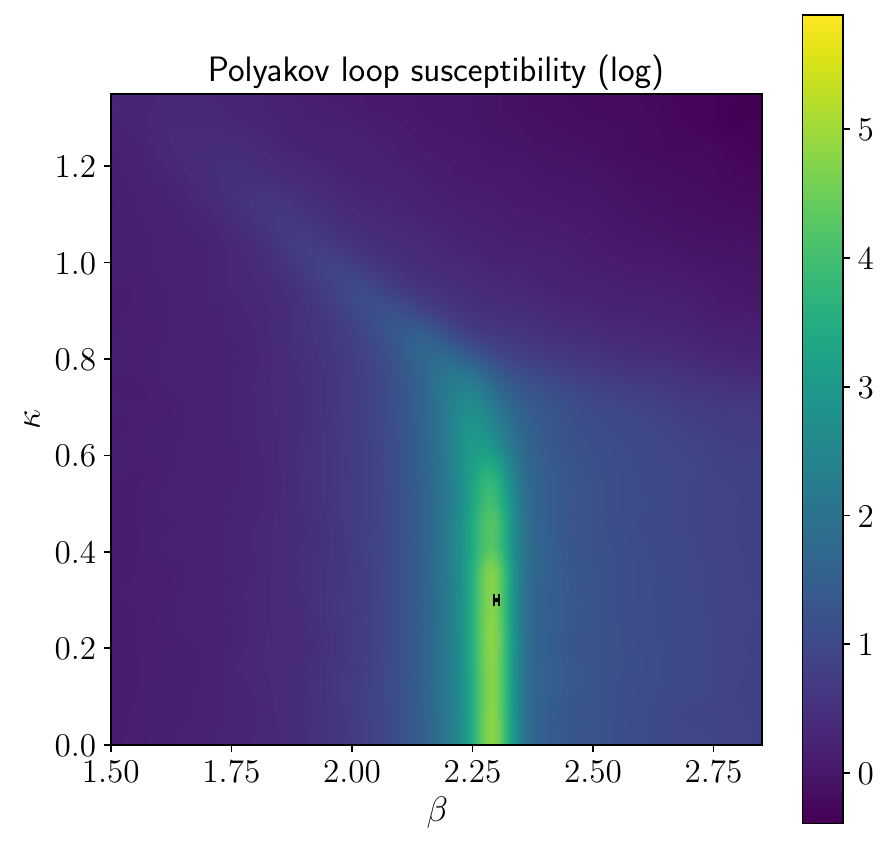}
    \caption{Heatmap plot of the susceptibility $\chi_G$ of the
      gauge-Higgs coupling term $G$ (top panel), the plaquette
      susceptibility $\chi_U$ (center panel), and the logarithm of the
      Polyakov-loop susceptibility $\chi_P$ (bottom panel), see
      Eqs.~\eqref{eq:su2h7} and \eqref{eq:su2h8}. Here $N_s=20$ and
      $N_t=4$. In the top panel, the black point shows where the
      mobility edge $\lambda_c=\lambda_c(\kappa)$ has an inflection
      point along the line at constant $\beta=2.6$, see
      Fig.~\ref{fig:32}. In the center panel, it shows where the
      mobility edge $\lambda_c=\lambda_c(\beta)$ vanishes along the
      line at constant $\kappa=1.0$, see Fig.~\ref{fig:31}. In the
      bottom panel, it shows where the mobility edge
      $\lambda_c=\lambda_c(\beta)$ vanishes along the line at constant
      $\kappa=0.3$, see Fig.~\ref{fig:30}.}
  \label{fig:2_0}
\end{figure}

\begin{figure}[tb]
  \centering
  \includegraphics[width=0.325\textwidth]{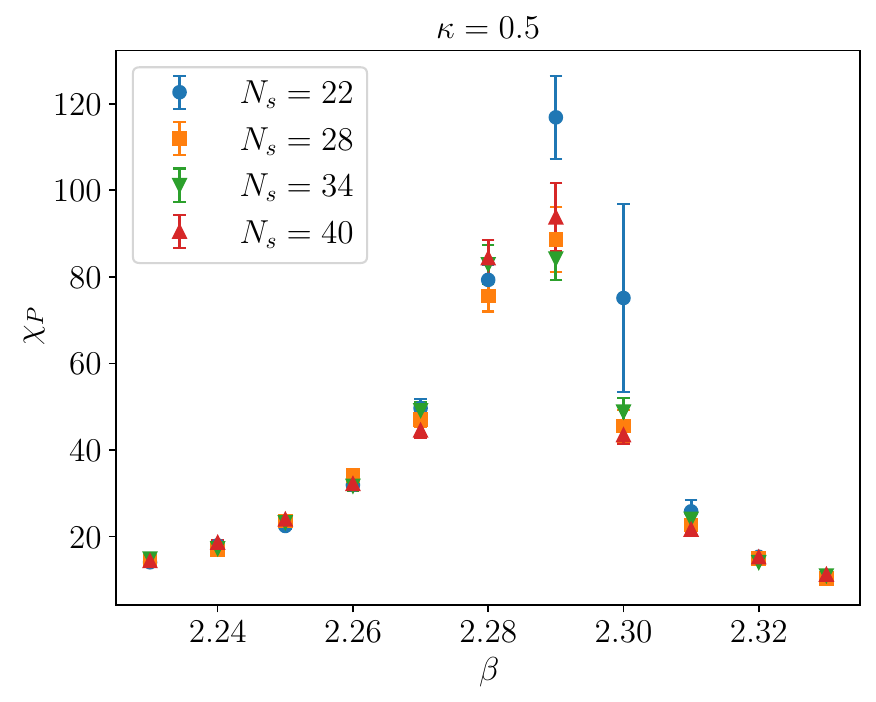}
  \hfil
  \includegraphics[width=0.325\textwidth]{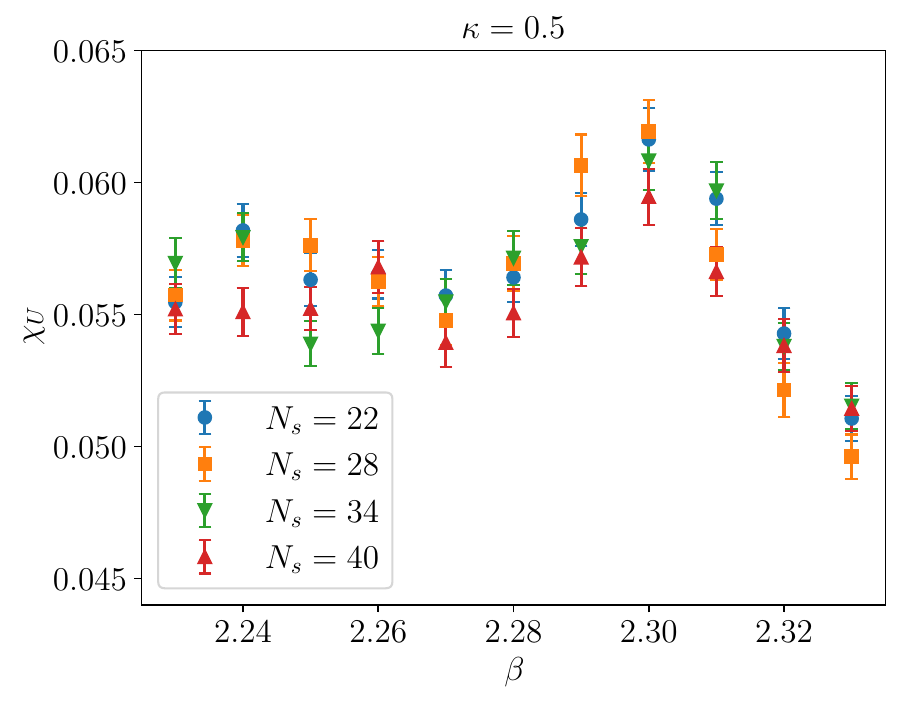}
  \hfil
  \includegraphics[width=0.325\textwidth]{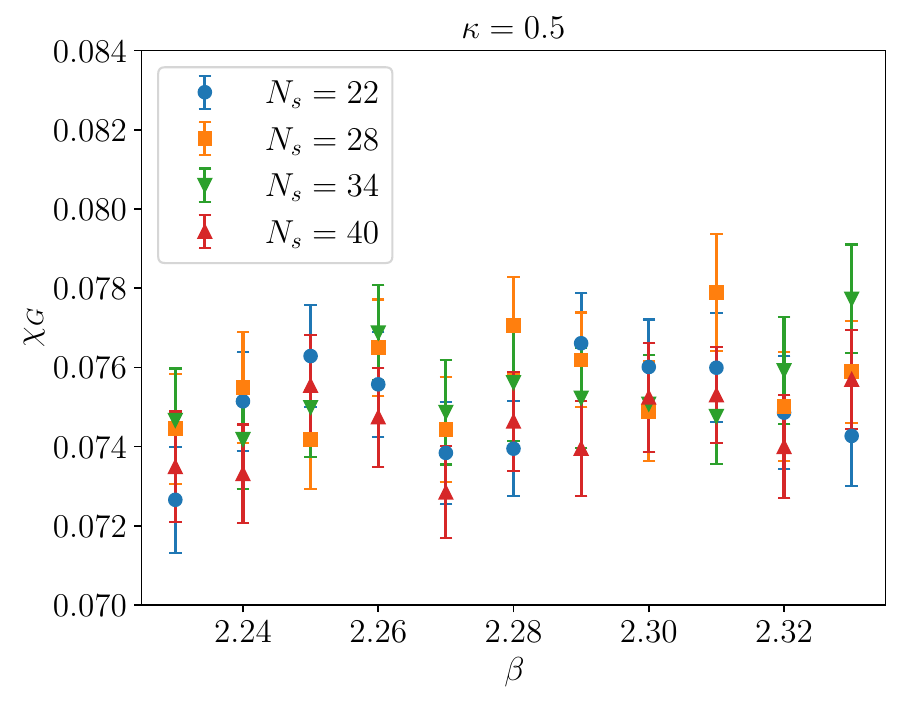}
  \caption{Polyakov-loop (top), plaquette (center), and gauge-Higgs
    coupling term (bottom) susceptibility across the transition
    between the confined and the deconfined phase at $\kappa=0.5$.
    Here $N_t=4$. The volume scaling is consistent with an analytic
    crossover.}
  \label{fig:13}
\end{figure}

\begin{figure}[tb]
  \centering
  \includegraphics[width=0.325\textwidth]{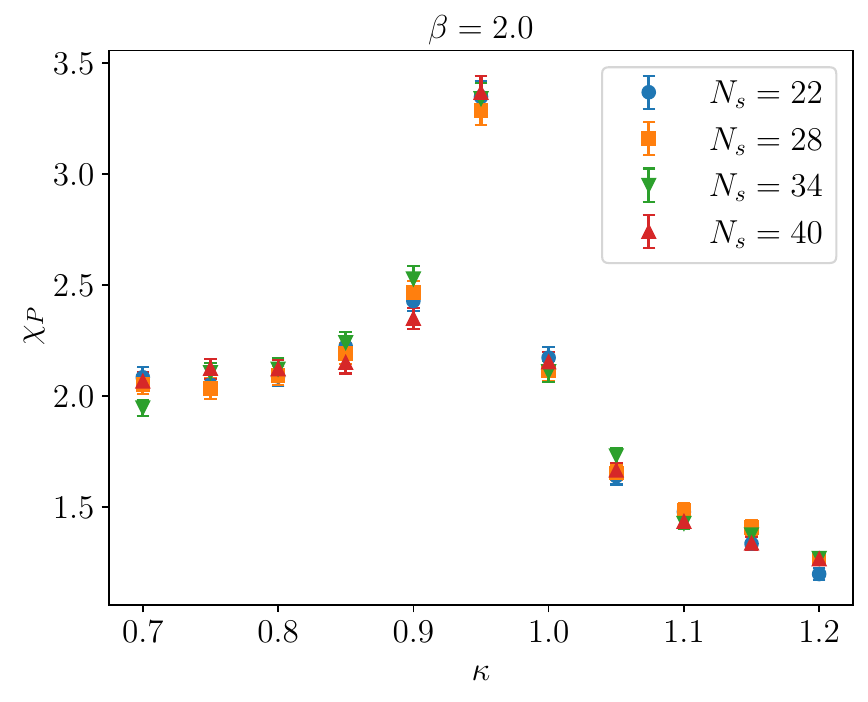}
  \includegraphics[width=0.325\textwidth]{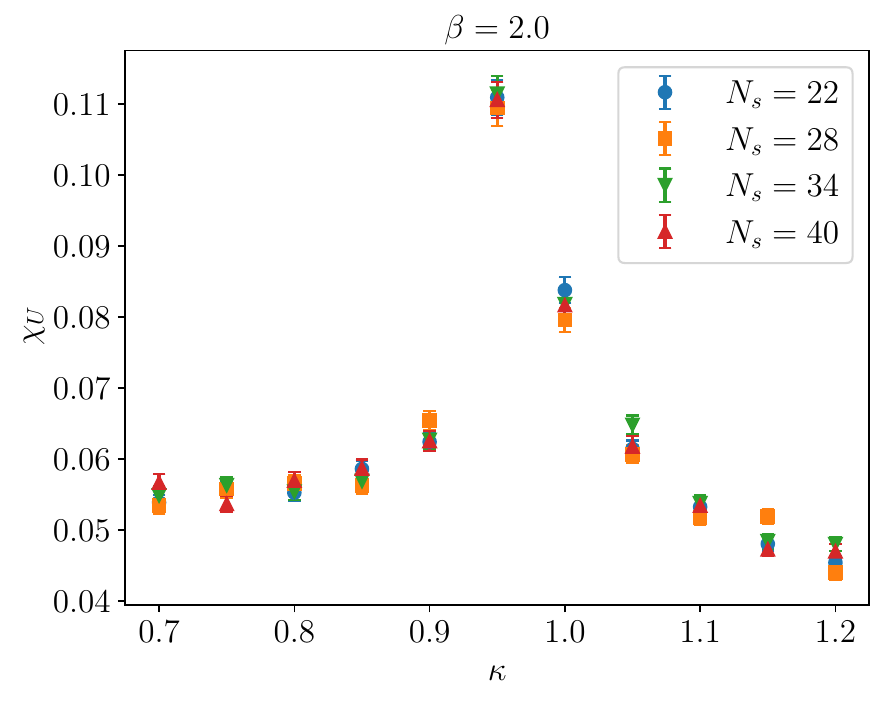}
  \includegraphics[width=0.325\textwidth]{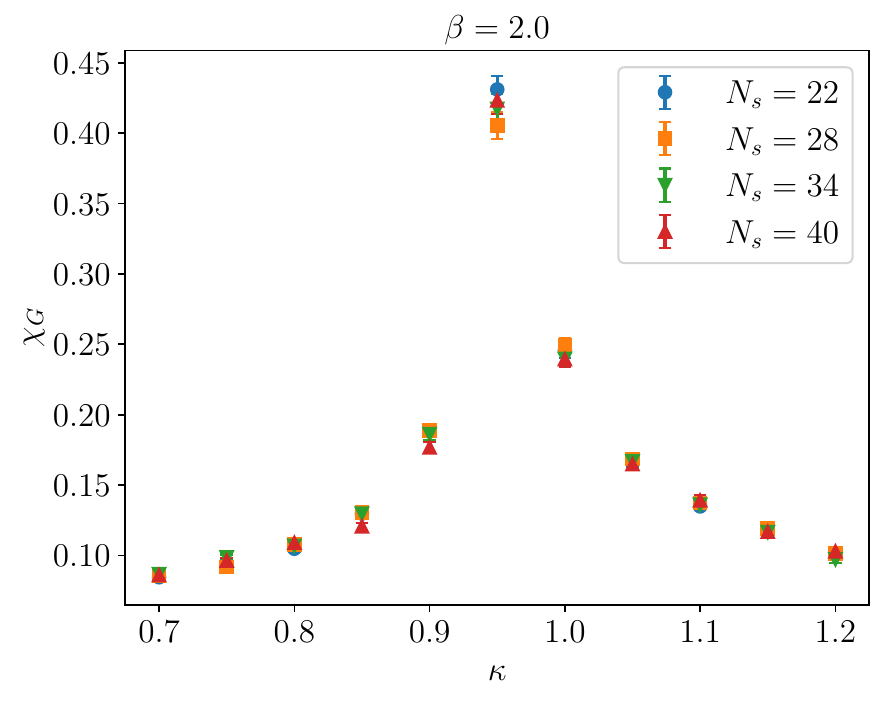}
  \caption{Polyakov-loop (top), plaquette (center), and gauge-Higgs
    coupling term (bottom) susceptibility near the transition between
    the confined and the Higgs phase at $\beta=2.0$. Here $N_t=4$. The
    volume scaling is consistent with an analytic crossover.}
  \label{fig:14_0}
\end{figure}

\begin{figure}[tb]
  \centering
  \includegraphics[width=0.325\textwidth]{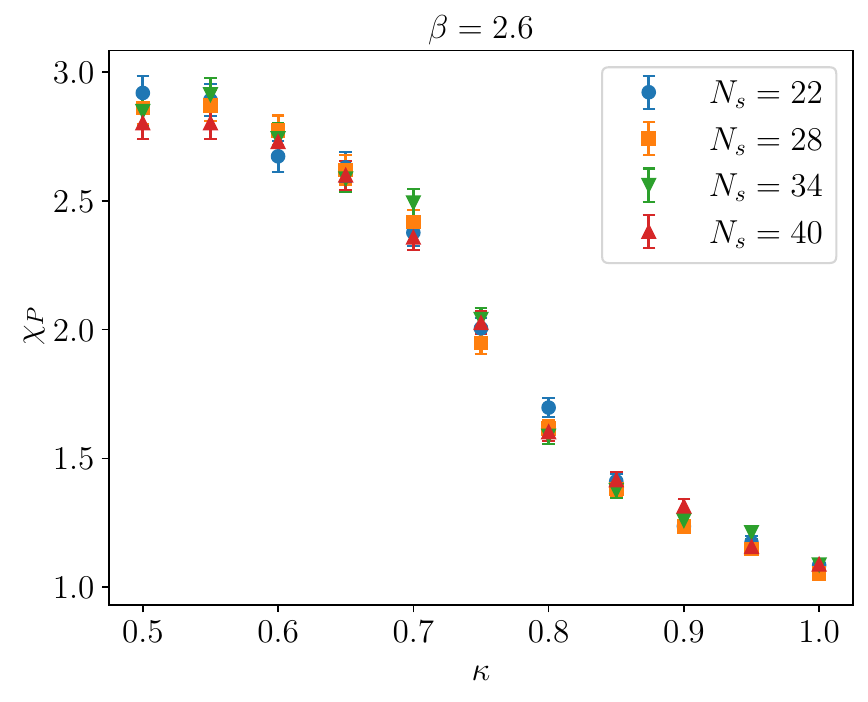}
  \includegraphics[width=0.325\textwidth]{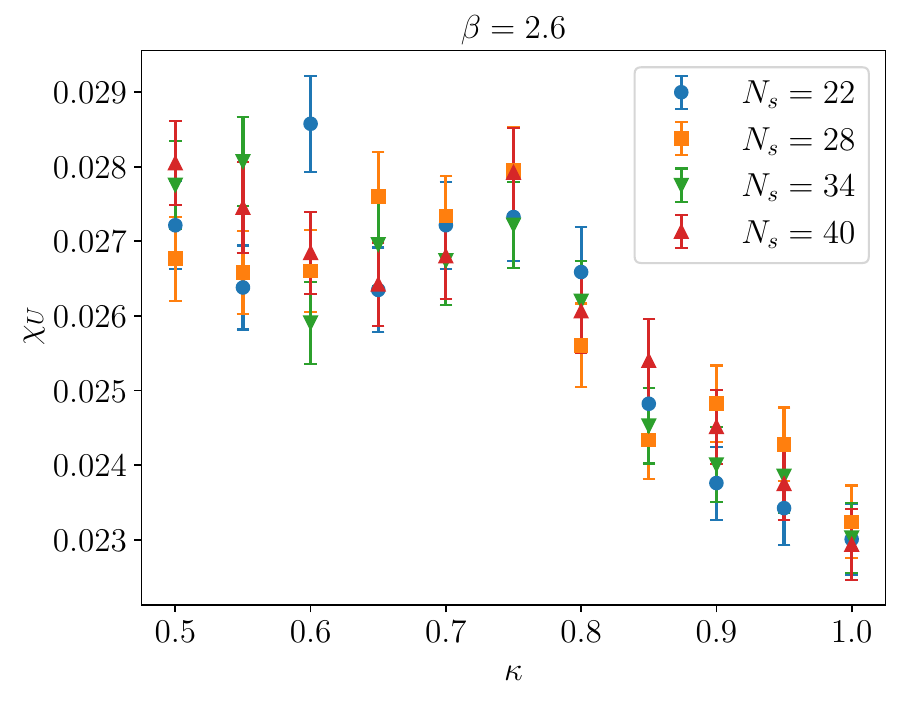}
  \includegraphics[width=0.325\textwidth]{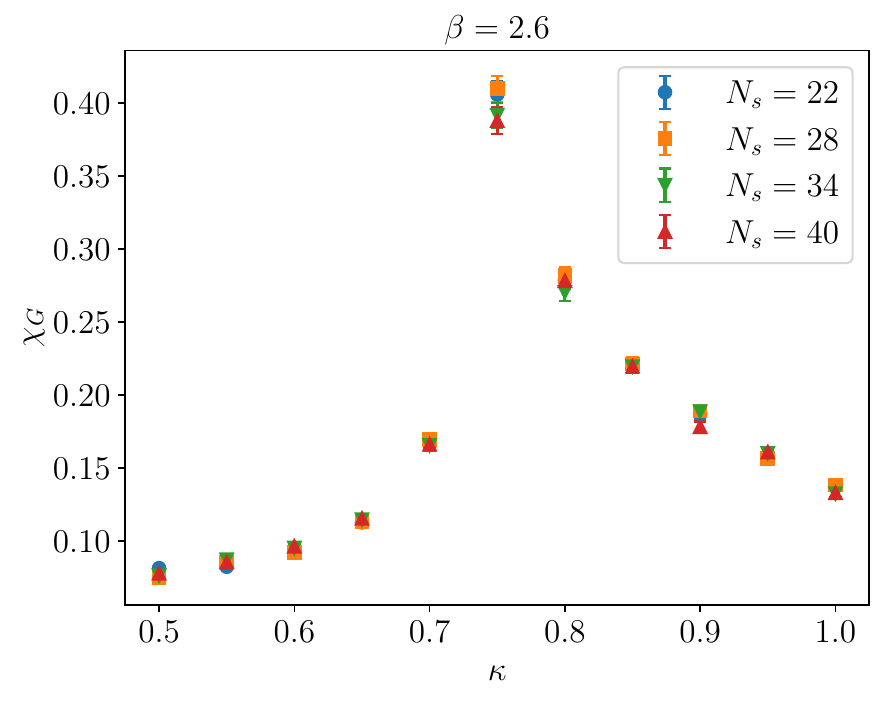}
  \caption{Polyakov-loop (top), plaquette (center), and gauge-Higgs
    coupling term (bottom) susceptibility near the transition between
    the deconfined and the Higgs phase at $\beta=2.6$. Here
    $N_t=4$. The volume scaling is consistent with an analytic
    crossover.}
  \label{fig:7}
\end{figure}

We show our results for the corresponding susceptibilities as heatmap
plots in Fig.~\ref{fig:2_0}.  Also in this case we estimated errors
(not shown in the figure) with a standard jackknife procedure, finding
them to be always within 3\%. In the top panel we show our results for
$\chi_G$. This quantity has a narrow ridge, visualized here as a
bright line, providing a clear separation between the Higgs phase and
the rest in most of the explored parameter space; a weakening of the
transition is visible in the top left part of the phase diagram.  In
the center panel we show the plaquette susceptibility $\chi_U$. This
separates clearly the confined phase from the Higgs phase, while the
ridge broadens at the transition between confined and deconfined phase
(as well as in the top left part of the phase diagram). In the bottom
panel we show the logarithm of the Polyakov-loop susceptibility. This
plot shows a bright line of strong transitions separating the confined
and deconfined phases. This line continues in the top left part of the
plot, still clearly separating the confined and Higgs phases, but it
is much dimmer there as the signal is two orders of magnitude weaker
than at the transition from the confined to the deconfined phase (see
Figs.~\ref{fig:13}, top and \ref{fig:14_0}, top). At the transition
between the deconfined and Higgs phase $\chi_P$ shows an inflection
point instead of a peak (see Fig.~\ref{fig:7}), with a sizeable
decrease in susceptibility corresponding here to a noticeable
darkening of the plot.

\begin{figure}[t]
  \centering
  \includegraphics[width=0.465\textwidth]{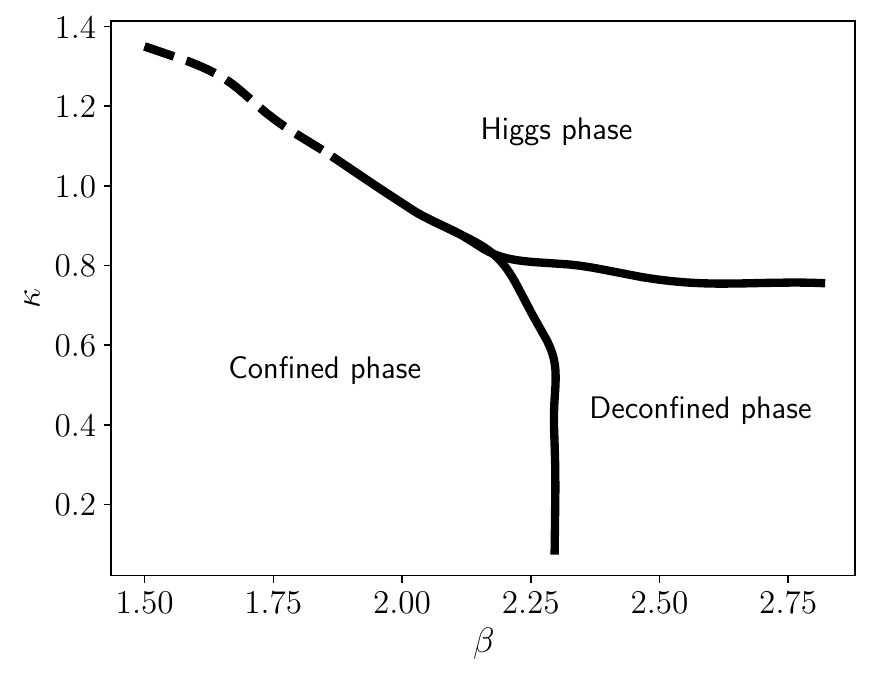}
  \caption{Schematic drawing of the phase diagram, obtained combining
    the maxima of the susceptibilities shown in Fig.~\ref{fig:2_0}.  A
    dashed line is used to indicate the weakening of the transition.}
  \label{fig:sketch}
\end{figure}

A sketch of the resulting phase diagram is shown in
Fig.~\ref{fig:sketch}, obtained by merging the various transition
lines, defined by the peaks of the suceptibilities. The dashed line at
low $\beta$ and large $\kappa$ signals a sizeable reduction in the
strength of the transition there, as shown by all three
observables. Except in this region, where they slightly deviate from
each other, the transition lines between confined and Higgs phase
obtained from the three different susceptibilities agree with each
other, so we drew a single line. Similarly, the transition lines
between confined and deconfined phase obtained from the plaquette and
the Polyakov loop susceptibility agree with each other, so we drew a
single line in this case as well.

To verify the expected crossover nature of the transitions, we studied
the volume dependence of the various susceptibilities in detail on
three lines, one at constant $\kappa=0.5$ and two at constant
$\beta=2.0$ and $\beta=2.6$, using lattices with
$N_s=22,28,34,40$. For each simulation point and each lattice volume
we used 4500 configurations.  We estimated errors by first averaging
over configurations in blocks of size $b_{\mathrm{size}}$ and
computing the standard jackknife error on the blocked ensemble, and
then increasing $b_{\mathrm{size}}$ until the error stabilized. For
our final estimates we used samples of size $b_{\mathrm{size}}=20$,
except at $\kappa=0.5$ where we used $b_{\mathrm{size}}=50$, although
this was really needed only around $\beta=2.3$.  We show our results
in Figs.~\ref{fig:13}--\ref{fig:7}.

In Fig.~\ref{fig:13} we show $\chi_P$, $\chi_U$ and $\chi_G$ along a
line of constant $\kappa=0.5$ across the transition from the confined
to the deconfined phase. The signal is very strong in $\chi_P$, and a
small peak is visible also in $\chi_U$. The location of these peaks is
not far from the critical point $\beta_c\approx 2.3$ of the pure gauge
theory at
$\kappa=0$~\cite{Engels:1989fz,Engels:1992fs,Engels:1992ke,Engels:1995em}.
The relatively large error bars found for $N_s=22$ between
$\beta=2.28$ and $\beta=2.31$, especially at $\beta=2.3$, are most
likely a finite-size effect due to the vicinity of the critical point
of the pure gauge theory, and are not observed on larger volumes. On
the other hand, no peak is visible in $\chi_G$, which is constant
within errors across the transition. This makes the gauge-Higgs
coupling term $G$ unsuitable to detect this transition.


In Fig.~\ref{fig:14_0} we show $\chi_P$, $\chi_U$ and $\chi_G$ along a
line of constant $\beta=2.0$ across the transition from the confined
to the Higgs phase. A clear peak is visible in all three observables,
with $\chi_P$ two orders of magnitude smaller than in
Fig.~\ref{fig:13} (top), and $\chi_U$ a factor of 2 larger than in
Fig.~\ref{fig:13} (bottom).

Finally, in Fig.~\ref{fig:7} we show $\chi_P$, $\chi_U$ and $\chi_G$
along a line of constant $\beta=2.6$ across the transition from the
deconfined to the Higgs phase. We observe a peak in $\chi_G$, of
similar magnitude as the one in Fig.~\ref{fig:14_0} (bottom) for the
transition from the confined to the Higgs phase. Neither $\chi_U$ nor
$\chi_P$ show any significant peak: $\chi_U$ changes slope at the
transition, while $\chi_P$ shows an inflection point.  This makes $U$
and $P$ not quite suitable observables to detect this transition.

While these results do not logically exclude the possibility of
genuine phase transitions at some points in the phase diagram,
combined with the results of Ref.~\cite{Bonati:2009pf} they make it
implausible.

\section{Localization properties of Dirac eigenmodes}
\label{sec:loc}

\begin{table}[t]
  \begin{tabular}[t]{|l|l|l|}
    \hline
    \multicolumn{3}{|c|}{$(\beta=1.9, \kappa=1.0)$} \\
    \hline
    $N_s$  & \#configurations & \#eigenvalues \\ \hline
    16 & 3000             & 33        \\ \hline
    20 & 1500             & 63             \\ \hline
  \end{tabular}

  \mbox{}
  
\begin{tabular}[t]{|l|l|l|}
  \hline
  \multicolumn{3}{|c|}{
  $(\beta=2.1, \kappa=1.0)$ and  $(\beta=2.6, \kappa=0.3)$} \\
  \hline
  $N_s$  & \#configurations & \#eigenvalues \\ \hline
  20 & 8970             & 63           \\ \hline
  24 & 8000             & 110             \\ \hline
  28 & 3000             & 174             \\ \hline
  32 & 1150             & 260           \\ \hline
\end{tabular}
\caption{Configuration statistics and number of (non-degenerate)
  eigenvalues used to study the volume scaling of the localization
  properties of staggered eigenmodes in the confined phase (top table)
  and in the deconfined and Higgs phases (bottom table).}
\label{table:2}
\end{table}

In this section we discuss the localization properties of the
eigenmodes of the staggered operator and how these correlate with the
gauge and Higgs fields, and we present a detailed test of the
sea/islands mechanism. We obtained the lowest modes of
$D^{\textrm{stag}}$ using the PRIMME
package~\cite{PRIMME,svds_software} for sparse matrices, exploiting
Chebyshev acceleration for faster convergence. The use of algorithms
for sparse matrices allows us to reduce the scaling of computational
time from $N_s^9$, expected for full diagonalization, down to $N_s^6$.

We first analyzed the eigenmodes in detail at three points of the phase
diagram, using several lattice volumes to study the scaling of
eigenvector and eigenvalue observables with the system size.  These
points are $\beta=1.9, \kappa=1.0$, in the confined phase, right below
the transition to the Higgs phase at constant $\kappa$
($\beta/\beta_c \approx 0.97$, with $\beta_c \approx 1.95$
corresponding to the peak in the Polyakov-loop susceptibility);
$\beta=2.1, \kappa=1.0$, in the Higgs phase, not far above the
transition between the two phases ($\beta/\beta_c \approx 1.08$); and
$\beta=2.6, \kappa=0.3$, deep in the deconfined phase.  We looked at
two lattice volumes in the confined phase, and at four lattice volumes
in the deconfined and Higgs phases; see Tab.~\ref{table:2} for details
about system size, configuration statistics, and number of eigenmodes.
We then computed the relevant observables locally in the spectrum,
approximating Eq.~\eqref{eq:su2h10} by averaging over spectral bins of
size $\Delta\lambda=0.0025$ at $\beta=1.9$, $\kappa=1.0$ (confined
phase), $\Delta\lambda= 0.01$ at $\beta=2.6$, $\kappa=0.3$ (deconfined
phase), and $\Delta\lambda= 0.0075$ at $\beta=2.1$, $\kappa=1.0$
(Higgs phase). Our results, reported in sections \ref{sec:loc_evec}
and \ref{sec:mobedge}, demonstrate low-mode localization in the
deconfined and in the Higgs phase.

This detailed study also allowed us to estimate the critical value of
$I_{s_0}$, which we could then use to efficiently determine the
dependence of the mobility edge, $\lambda_c$, on the parameters
$\beta$ and $\kappa$. We did this on two lines at constant $\kappa$:
one in the deconfined phase with $\kappa=0.3$, changing $\beta$ in the
interval $[2.35,2.60]$ with increments $\Delta\beta=0.05$; and one in
the Higgs phase with $\kappa=1.0$, changing $\beta$ in $[2.1,2.4]$
with increments $\Delta\beta=0.05$. We also studied one line at
constant $\beta=2.6$, changing $\kappa$ in $[0.35,1.0]$ in increments
of $\Delta\kappa=0.05$. Here we used a single volume ($N_s=20, N_t=4$)
and 3000 configurations at each point (except for the three points
already discussed above). In all these calculations we computed
$I_{s_0}$ locally in the spectrum averaging over bins of size
$\Delta\lambda=0.008$. Our results, reported in section
\ref{sec:mobedgbk}, show that along both lines at constant $\kappa$
the mobility edge disappears at a critical $\beta$ near the crossover
to the confined phase; and that along the line at constant $\beta$ the
mobility edge is always nonzero, but it changes behavior at the
crossover between the deconfined and the Higgs phase.

We then studied the correlation between localized modes and the
fluctuations of the gauge and Higgs fields, and tested the refined
sea/islands picture of Ref.~\cite{Baranka:2022dib}. Our results,
reported in section \ref{sec:bos}, show a strong correlation with
Polyakov-loop and plaquette fluctuations, and an even stronger
correlation with the fluctuations identified in
Ref.~\cite{Baranka:2022dib} as the most relevant to localization.

\subsection{Eigenvector observables}
\label{sec:loc_evec}

\begin{figure}[t]
  \centering
  \includegraphics[width=0.45\textwidth]{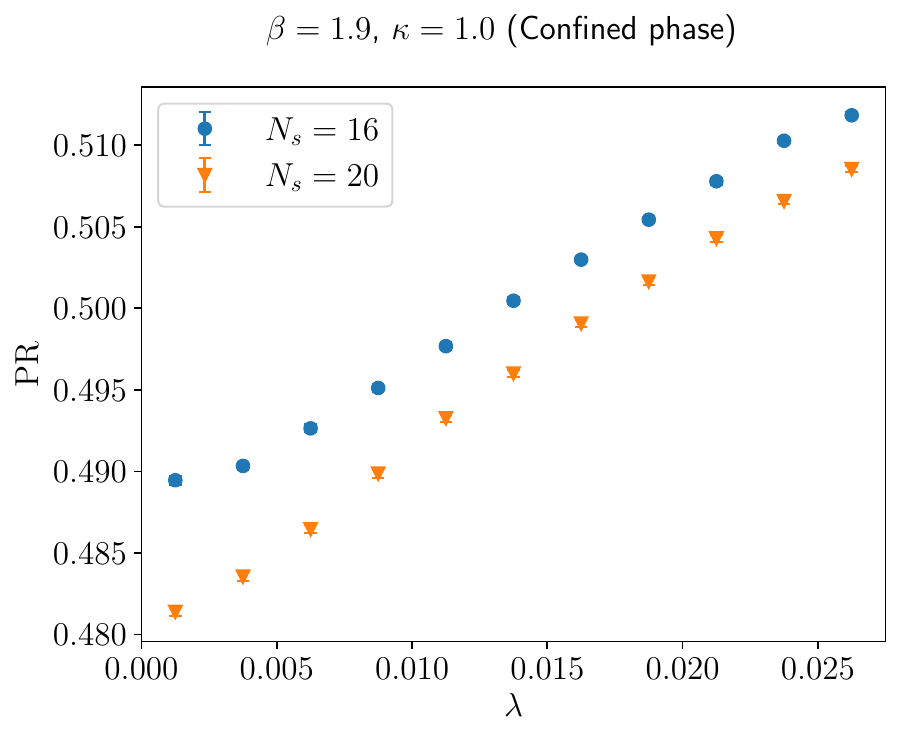}\hfil
  \includegraphics[width=0.45\textwidth]{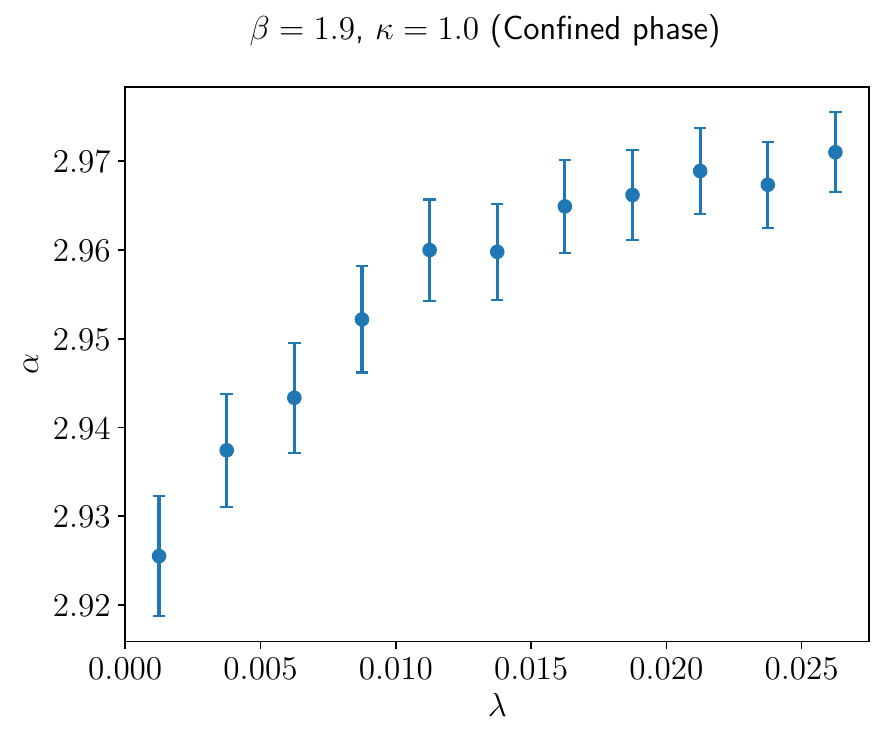}
  \caption{Participation ratio, Eq.~\eqref{eq:su2h9}, of the low
    staggered eigenmodes at $\beta=1.9$ and $\kappa=1.0$ in the
    confined phase for two different spatial volumes (top panel) and
    corresponding fractal dimension estimated using
    Eq.~\eqref{eq:fracdim2} with $N_{s_1}=16$, $N_{s_2}=20$ (bottom
    panel). Here $N_t=4$.}
  \label{fig:16}
\end{figure}
  
In the top panel of Fig.~\ref{fig:16} we show the PR of the modes in
the confined phase.  The PR is slightly larger for $N_s=16$ than for
$N_s=20$, signaling that the fractal dimension is smaller than 3. This
is shown explicitly in the bottom panel, where we plot
$\alpha(\lambda)$, see Eq.~\eqref{eq:fracdim}. This is estimated
numerically from a pair of volumes as
\begin{equation}
  \label{eq:fracdim2}
  \alpha_{\mathrm{num}}(\lambda;N_{s1},N_{s2}) = 3+
  \f{\log
    \f{\mathrm{PR}(\lambda,N_{s1})}{\mathrm{PR}(\lambda,N_{s2})}}
  {\log\f{N_{s1}}{N_{s2}}}\,.
\end{equation}
The fractal dimension of near-zero modes is slightly below 3, and
approaches 3 as one moves up in the spectrum. Taken at face value,
this means that these modes are only slightly short of being fully
delocalized. Clearly, this effect could be just a finite-size artifact
due to the small volumes employed here. However, it could also signal
that a ``geometric'' transition is approaching, where a mobility edge
and, correspondingly, critical modes appear at the origin.

\begin{figure}[t]
  \centering
  \includegraphics[width=0.45\textwidth]{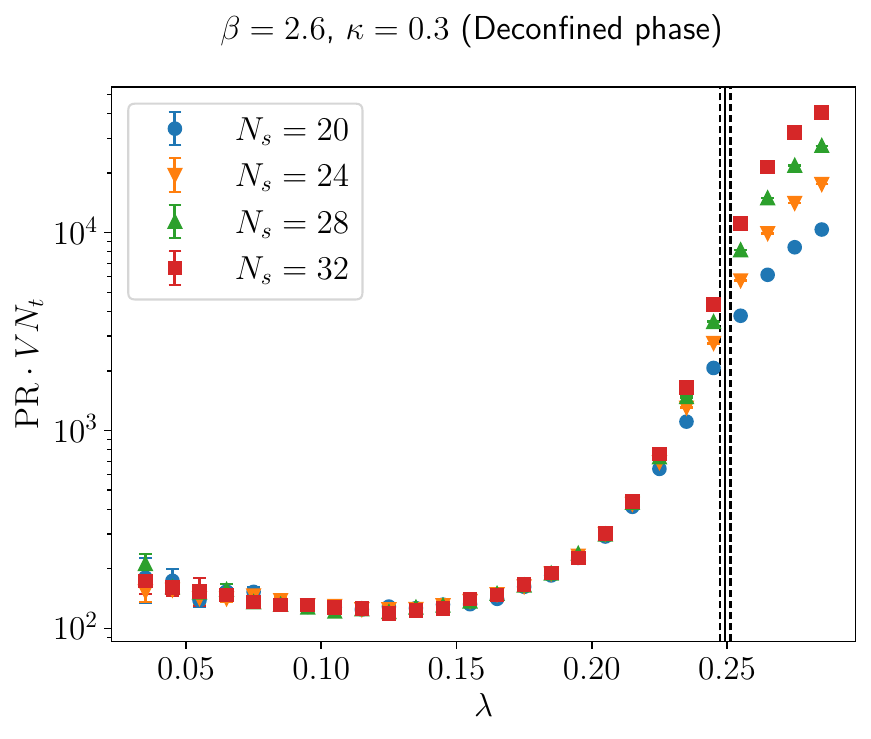}\hfil
  \includegraphics[width=0.45\textwidth]{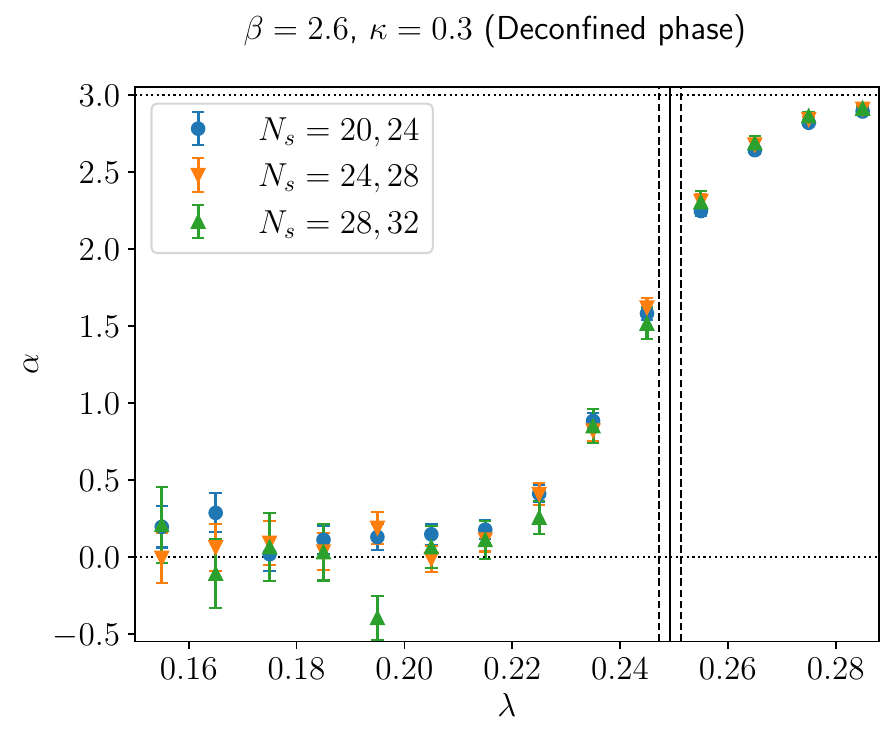}
  
  \caption{The mode size $N_tV\cdot \mathrm{PR}= \mathrm{IPR}^{-1}$,
    Eq.~\eqref{eq:su2h9}, of the low staggered eigenmodes for
    different volumes (top panel), and corresponding fractal
    dimension estimated using Eq.~\eqref{eq:fracdim2} with three
    different volume pairs (bottom panel), at $\beta=2.6$ and
    $\kappa=0.3$ in the deconfined phase.  Here $N_t=4$. The vertical
    solid line shows the position of the mobility edge, vertical
    dashed lines indicate the corresponding error band. In the bottom
    panel, horizontal dotted lines mark the values $\alpha=0$,
    corresponding to localized modes, and $\alpha=3$, corresponding to
    totally delocalized modes.}
  \label{fig:20}
\end{figure}

\begin{figure}[t!]
  \centering
  \includegraphics[width=0.45\textwidth]{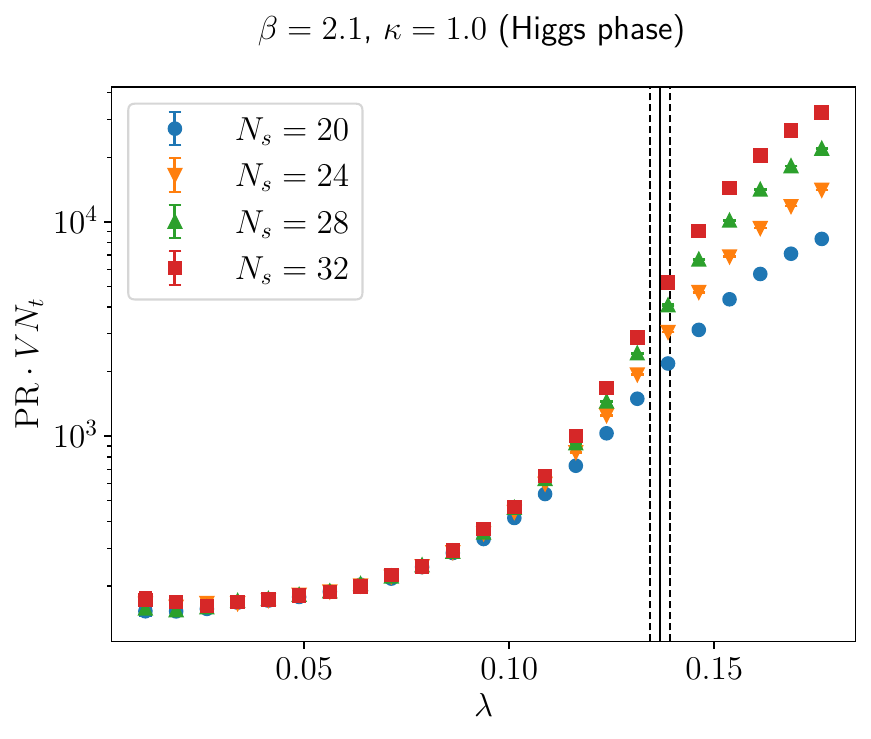}\hfil
    \includegraphics[width=0.45\textwidth]{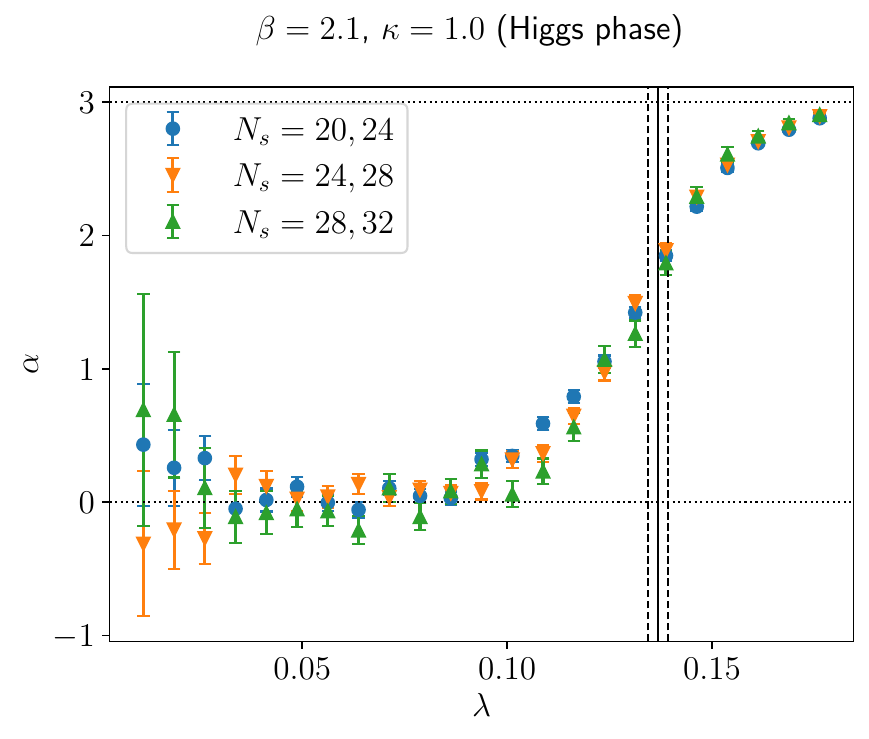}
    \caption{As in Fig.~\ref{fig:20}, but at $\beta=2.1$ and
      $\kappa=1.0$ in the Higgs phase.}
  \label{fig:22}
\end{figure}

\begin{figure}[t!]
  \centering
  \includegraphics[width=0.45\textwidth]{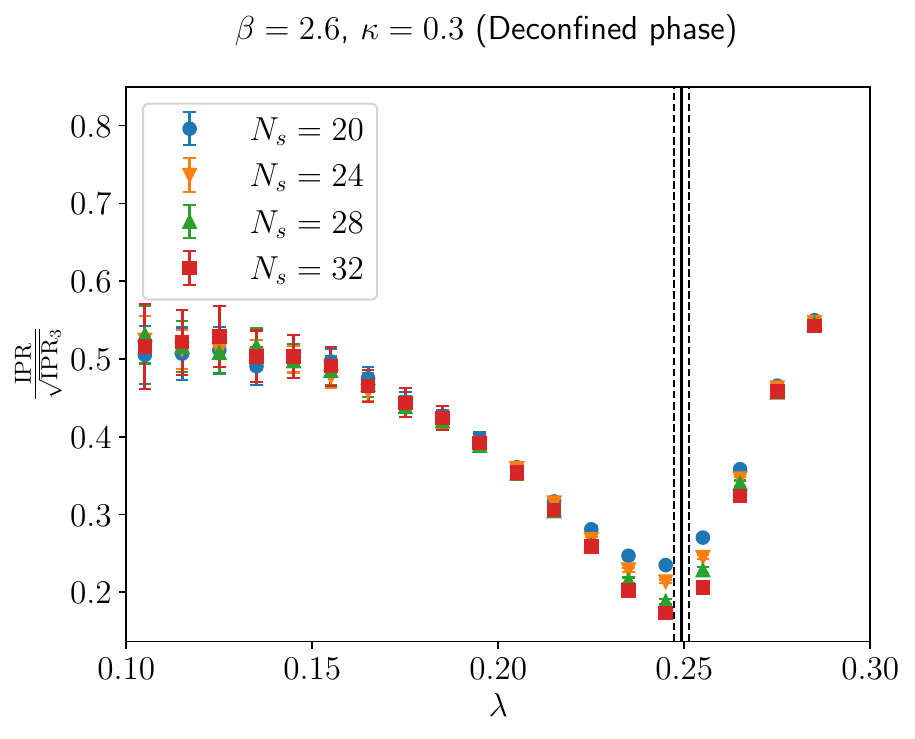}\hfil
  \includegraphics[width=0.45\textwidth]{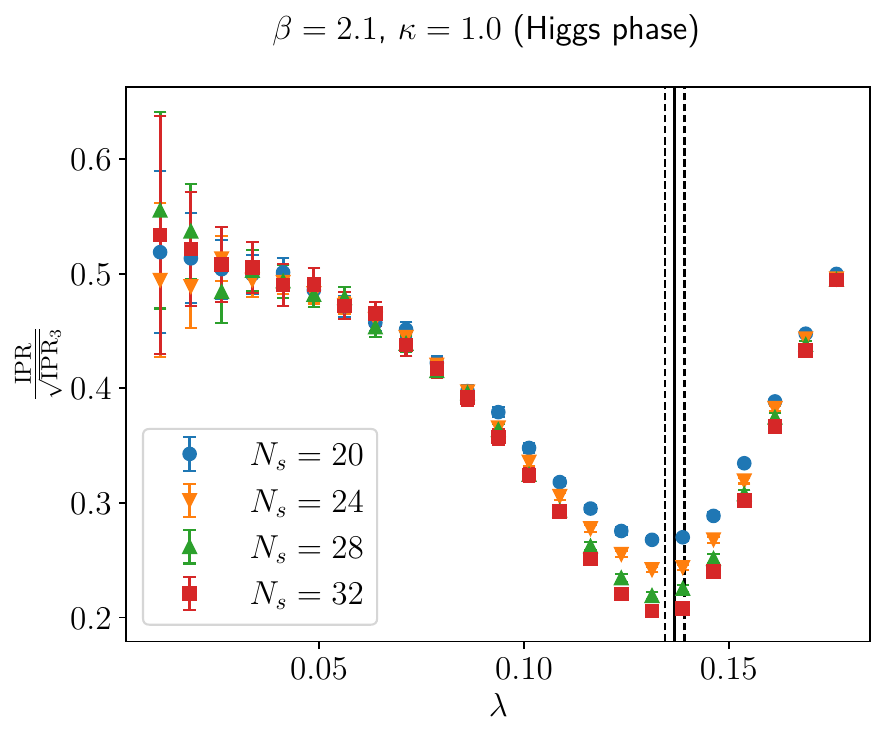}

  \caption{Ratio of generalized IPRs, Eq.~\eqref{eq:iprratio}, at
    $\beta=2.6$ and $\kappa=0.3$ in the deconfined phase (top panel),
    and at $\beta=2.1$ and $\kappa=1.0$ in the Higgs phase (bottom
    panel). The vertical solid line shows the position of the mobility
    edge, vertical dashed lines indicate the corresponding error
    band. A nontrivial volume scaling indicates nontrivial
    multifractal properties of the eigenmodes at the mobility edge.}
  \label{fig:new4}
\end{figure}

In the top panels of Figs.~\ref{fig:20} and \ref{fig:22} we show the
size $N_t V\cdot \mathrm{PR} = \mathrm{IPR}^{-1}$ of the modes in the
deconfined and in the Higgs phase, respectively. In both cases the
size of the lowest modes does not change with the volume, showing that
they are localized.  Higher up towards the bulk of the spectrum the
mode size shows a strong volume dependence. Above a certain point in
the spectrum this is compatible with a linear scaling in the volume,
indicating that these modes are delocalized. The point where this
starts to happen is consistent with the mobility edge, determined
below in section \ref{sec:mobedge} using spectral statistics, and
marked in these plots by a solid vertical line (with an error band
shown with dashed lines). The localization properties of low and bulk
modes in the deconfined and in the Higgs phase are made quantitative
in the bottom panels of Figs.~\ref{fig:20} and \ref{fig:22}, where we
show their fractal dimension. For low modes this is zero within
errors. Near the mobility edge our estimates for $\alpha$ increase
towards 3, which they almost reach at the upper end of the available
spectral range. The rise should become steeper when using pairs of
larger volumes, leading to a jump from 0 to 3 at the mobility edge in
the infinite-volume limit. Such a tendency is visible in the Higgs
phase. Our results are also consistent with modes at the mobility edge
displaying critical localization properties, with a fractal dimension
between 1 and 2.

The nontrivial multifractal properties of the eigenmodes at the
mobility edge are made evident in Fig.~\ref{fig:new4}, where we show
the ratio
\begin{equation}
  \label{eq:iprratio}
  \f{\mathrm{IPR}_2(\lambda,N_s)}{\sqrt{\mathrm{IPR}_3(\lambda,N_s)}}
  \sim N_s^{-\left(D_2(\lambda)-D_3(\lambda)\right)}\,,
\end{equation}
where the generalized IPRs have been defined in Eq.~\eqref{eq:genIPR}.
This quantity tends to a constant both in the localized ($D_q=0$) and
in the delocalized regime ($D_q=3$), while it has a nontrivial volume
scaling for modes displaying multifractality, i.e, with $q$-dependent
$D_q$. This is expected to be a feature of the critical modes found at
the mobility edge. This point in the spectrum is indeed characterized
by a nontrivial volume scaling of the ratio in
Eq.~\eqref{eq:iprratio}, which also reaches its minimum in the
vicinity of the mobility edge.

Comparing results in the confined and in the Higgs phase, that lie on
the same line at constant $\kappa$ near the transition, one sees that
the rapid change in the localization properties of the low modes takes
place precisely in the crossover region. This issue is studied in more
detail below in section \ref{sec:mobedgbk}.

\subsection{Eigenvalue observables and mobility edge}
\label{sec:mobedge}

\begin{figure}[t]
  \centering
  \includegraphics[width=0.4\textwidth]{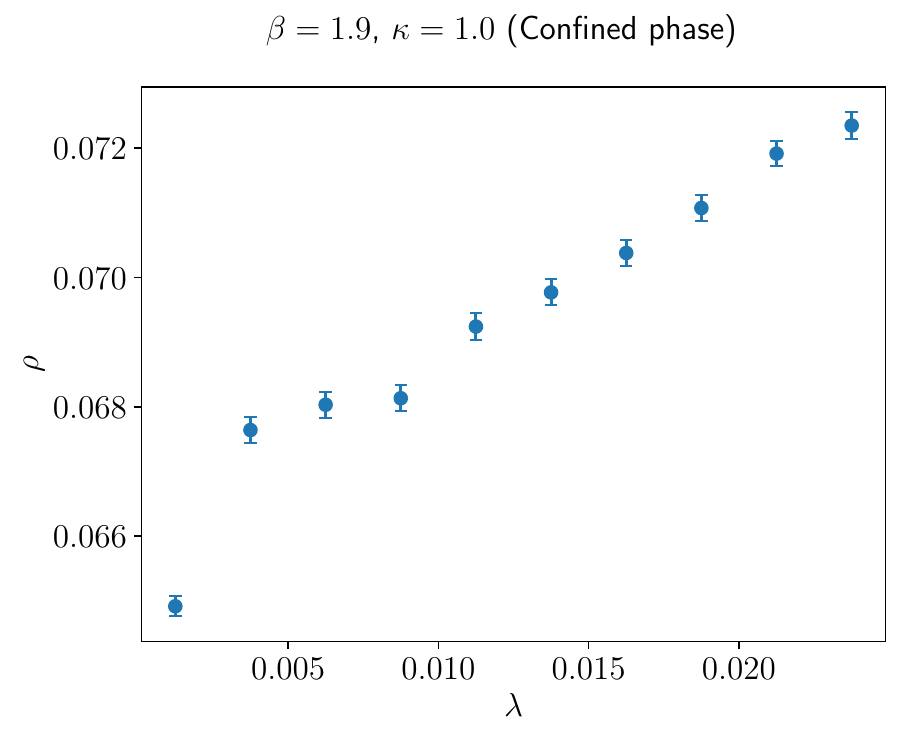}

  \includegraphics[width=0.4\textwidth]{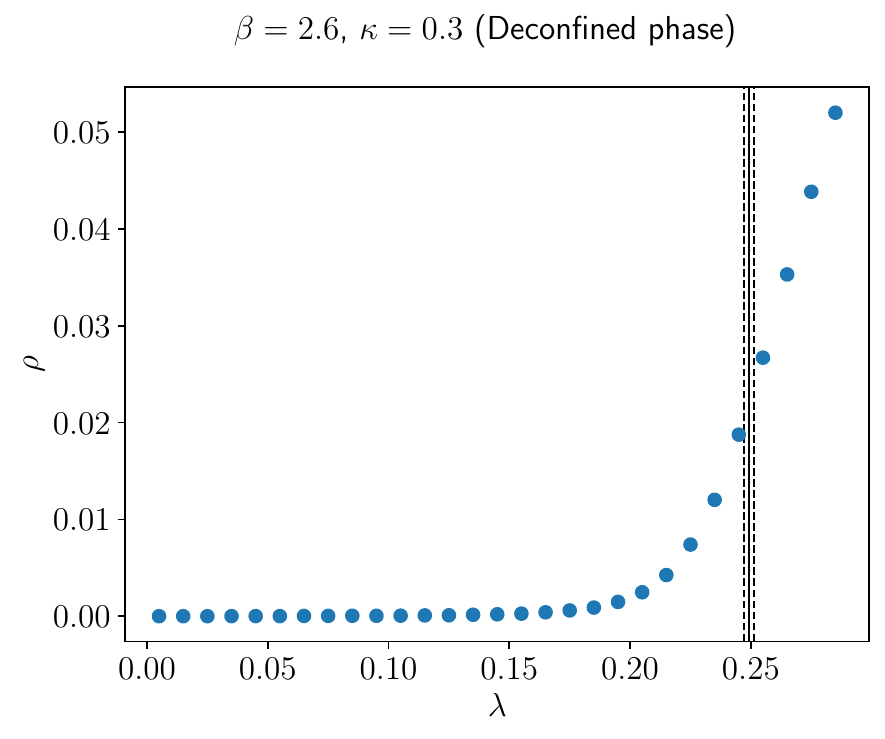}
  
  \includegraphics[width=0.4\textwidth]{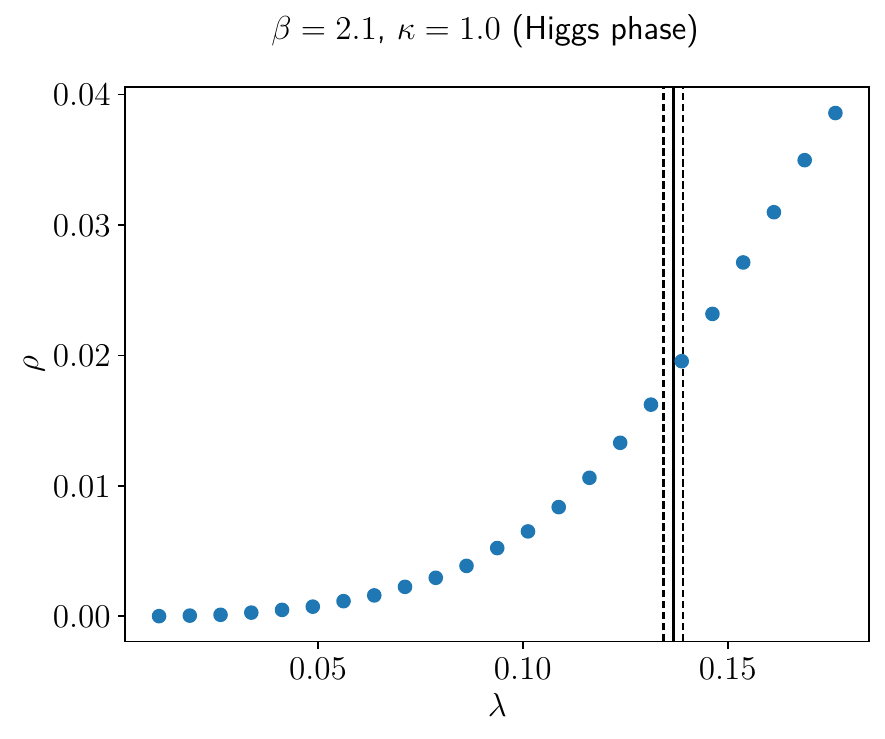}
  \caption{The spectral density at $\beta=1.9$ and $\kappa=1.0$ in the
    confined phase (top panel; here $N_s=20$), at $\beta=2.6$ and
    $\kappa=0.3$ in the deconfined phase (center panel; here
    $N_s=32$), and at $\beta=2.1$ and $\kappa=1.0$ in the Higgs phase
    (bottom panel; here $N_s=32$). In all plots $N_t=4$.}
  \label{fig:38}
\end{figure}
  
We now discuss eigenvalue observables, starting from the spectral
density, Eq.~\eqref{eq:su2h13}, shown in Fig.~\ref{fig:38}. In the
confined phase (top panel) the spectral density is practically
constant in the lowest bins (except for the very lowest, which is
depleted due to the smallness of the lattice volume), and grows as one
moves towards the bulk of the spectrum. If we were in the chiral limit
of massless fermions, a nonzero spectral density near the origin would
indicate the spontaneous breaking of chiral
symmetry~\cite{Banks:1979yr}. Being in the opposite limit of
infinitely massive fermions, we can speak of spontaneous chiral
symmetry breaking only in a loose sense. In the deconfined and in the
Higgs phase (bottom panels) we see instead that the spectral density
is close to zero for near-zero modes, corresponding (again, loosely
speaking) to the restoration of chiral symmetry. As we increase
$\lambda$ the spectral density increases, and does so faster as one
approaches the mobility edge. However, no sign of critical behavior is
visible along the spectrum.

\begin{figure}[t]
  \centering
  \includegraphics[width=0.45\textwidth]{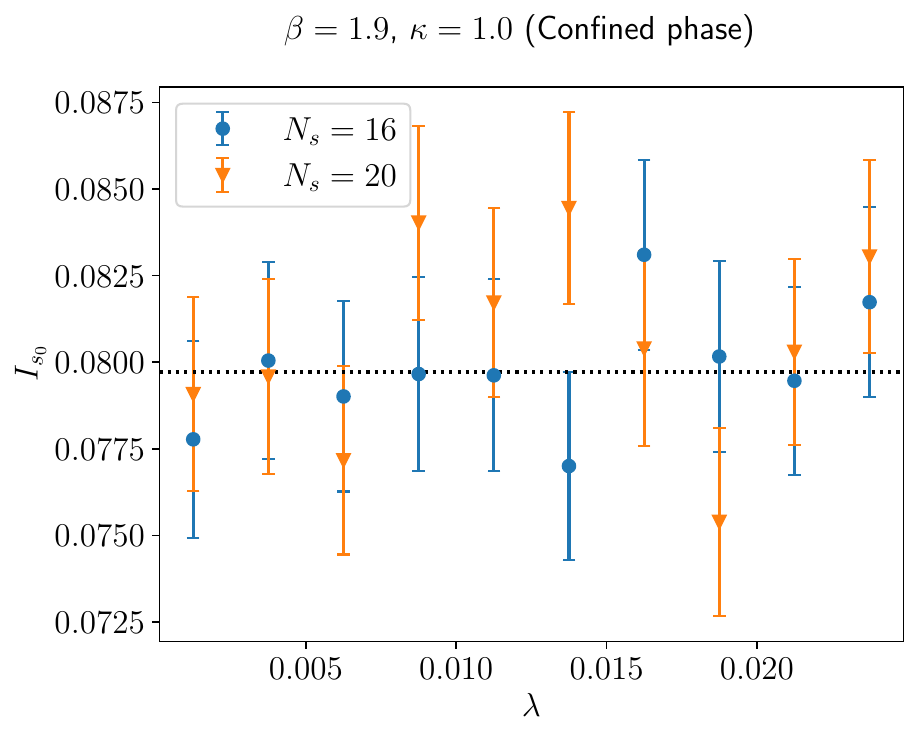}
  \caption{The integrated unfolded level spacing distribution
    $I_{s_0}$, Eq.~\eqref{eq:su2h14}, at $\beta=1.9$ and $\kappa=1.0$
    in the confined phase. Here $N_t=4$. The horizontal dotted line
    shows the value of $I_{s_0}$ expected for RMT statistics.}
  \label{fig:29}
\end{figure}

\begin{figure}[t]
  \centering
  \includegraphics[width=0.45\textwidth]{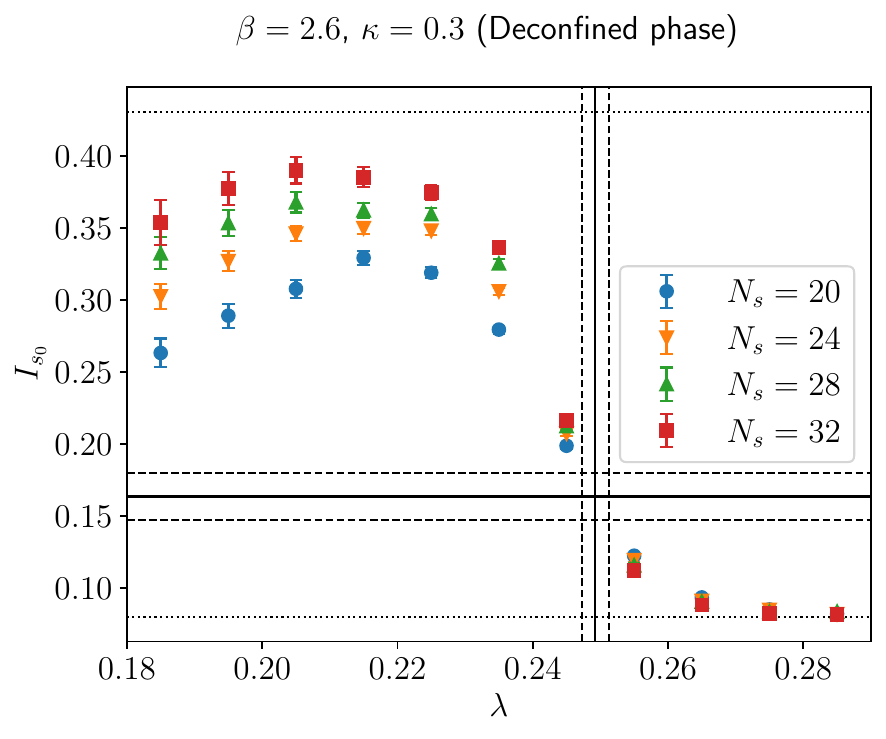}
  \caption{The integrated unfolded level spacing distribution
    $I_{s_0}$, Eq.~\eqref{eq:su2h14}, at $\beta=2.6$ and $\kappa=0.3$
    in the deconfined phase. Here $N_t=4$. The upper and lower
    horizontal dotted lines show the value of $I_{s_0}$ expected for
    Poisson statistics and for RMT statistics, respectively. The
    vertical solid and dashed lines indicate the position and the
    error band of the mobility edge. The horizontal solid and dashed
    lines correspond to the estimate for the critical value
    $I_{s_0,c}$ of $I_{s_0}$ at the mobility edge and its error band.}
  \label{fig:27}
\end{figure}
\begin{figure}[th!]
  \centering
  \includegraphics[width=0.45\textwidth]{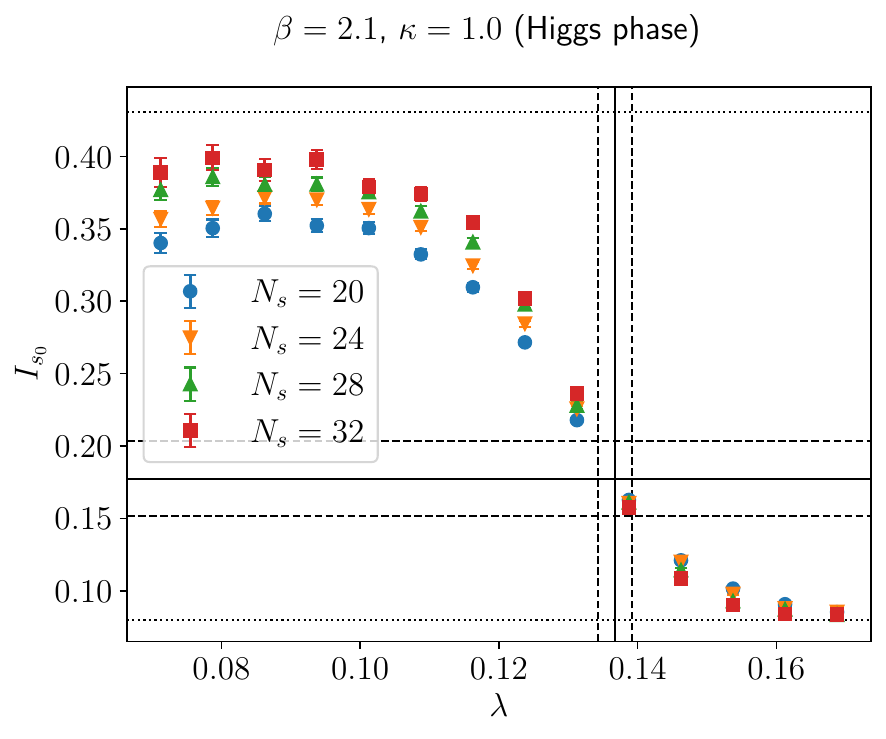}
  \caption{As in Fig.~\ref{fig:27}, but at $\beta=2.1$ and
    $\kappa=1.0$ in the Higgs phase. Here $N_t=4$.}
  \label{fig:28}
\end{figure}

We now move on to discuss the spectral statistic $I_{s_0}$,
Eq.~\eqref{eq:su2h14}, for the low modes in the three different phases
of the system. To estimate this quantity numerically we unfolded the
spectrum, averaging then $I_{s_0}$ in small spectral bins and over
gauge configurations. More precisely, we defined the unfolded spacings
using Eq.~\eqref{eq:su2h11}, using for
$\la\lambda_{l+1}-\lambda_l\ra_\lambda$ the average level spacing
found in a given spectral bin, including all pairs of eigenvalues for
which the smaller one fell in the bin.

In Fig.~\ref{fig:29} we show $I_{s_0}$ in the confined phase. As
expected, $I_{s_0}$ is compatible with the value predicted by RMT in
the whole available spectral range for both volumes, further
confirming that these modes are delocalized. In Figs.~\ref{fig:27} and
\ref{fig:28} we show the value of $I_{s_0}$ in the deconfined and in
the Higgs phase. For modes near the origin $I_{s_0}$ approaches the
value expected for Poisson statistics as we increase the volume,
signaling that these are localized modes. For higher modes the value
of $I_{s_0}$ tends instead to the RMT prediction as the volume
increases, showing that modes are delocalized in this spectral region.
Between these two regimes, we can find the mobility edge $\lambda_c$
as the point where $I_{s_0}$ is scale-invariant and the curves cross
each other.

To find $\lambda_c$ and the critical value $I_{s_0,c}$ of the spectral
statistic we interpolated the numerical data with natural cubic
splines, and determined the crossing point for the various pairs of
system sizes using Cardano's formula. The statistical error on each
determination of $\lambda_c$ and $I_{s_0,c}$ originating in the
numerical uncertainty on $I_{s_0}$ in the various bins is estimated by
obtaining the interpolating splines and their crossing point for a set
of synthetic data, generating 100 data sets by drawing for each bin a
number from a Gaussian distribution with mean equal to the average
$I_{s_0}$ in the bin and variance equal to the square of the
corresponding error. The systematic errors on $\lambda_c$ and
$I_{s_0,c}$ due to finite-size effects are estimated as the variance
of the set of values for the crossing point and corresponding value of
$I_{s_0}$ obtained from all the pairs of volumes. We finally estimated
the mobility edge and the critical $I_{s_0}$ as those obtained from
the crossing point of the biggest volume pair ($N_s=28,32$), as it
should be the closest to the actual value in the infinite-volume
limit, and the corresponding error by adding quadratically its
statistical error with the finite-size systematic error discussed
above. The total error is largely dominated by the finite-size
contribution. We did this separately for the configurations in the
deconfined and in the Higgs phase. The results for $\lambda_c$ and
$I_{s_0,c}$ are reported in Tab.~\ref{tab:is0c}, and shown in
Figs.~\ref{fig:27} and \ref{fig:28} as solid lines, with dashed lines
marking the corresponding error bands. The two determinations of
$I_{s_0,c}$, obtained in the deconfined and in the Higgs phase, agree
within errors. Despite the uncertainty on $I_{s_0,c}$ being 10--15\%,
we could determine $\lambda_c$ with a 1--2\% uncertainty thanks to the
steepness of $I_{s_0}$ near the mobility edge.

\begin{table}[b]
  \centering
  \begin{tabular}{cc|c|cc}
    $ \beta $&$ \kappa$& phase & $\lambda_c$ & $I_{s_0,c}$\\ \hline\hline
    $ 2.6 $ & $ 0.3$ & deconfined & 0.2493(20) & 0.164(16)\\
    $ 2.1 $ & $ 1.0 $  & Higgs &  0.1367(24)   & 0.177(26)\\ \hline
  \end{tabular}
  \caption{Mobility edge and critical value of $I_{s_0}$ estimated at
    two points of the phase diagram, one in the deconfined and one in
    the Higgs phase.}
  \label{tab:is0c}
\end{table}

\subsection{$\beta$ and $\kappa$ dependence of the mobility edge}
\label{sec:mobedgbk}

\begin{figure}[thb]
  \centering
  \includegraphics[width=0.45\textwidth]{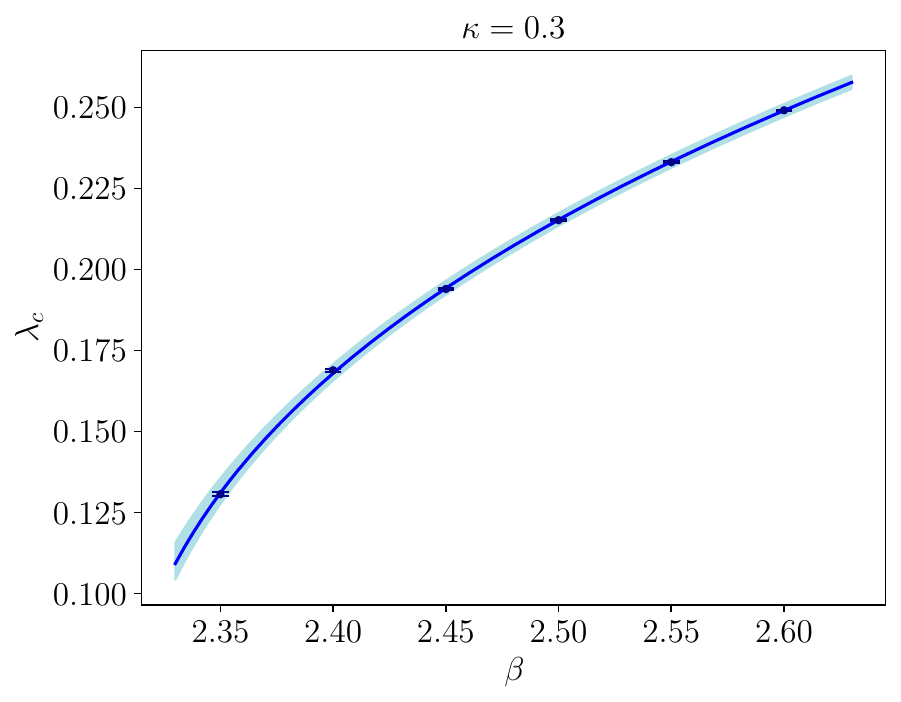}
  \caption{The dependence of the mobility edge on $\beta$ in the
    deconfined phase on the line at constant $\kappa=0.3$. The solid
    line is a power-law fit, Eq.~\eqref{eq:plaw}, to the numerical
    data; the band corresponds to the finite-size systematic
    uncertainty discussed in the text. The point where the mobility
    edge vanishes is estimated at $\beta_{\mathrm{loc}}=2.2997(57)$,
    in the crossover region between the confined and deconfined
    phases, see Fig.~\ref{fig:2_0} (bottom).}
  \label{fig:30}
\end{figure}

Having obtained estimates of $I_{s_0,c}$ we can now use them to
efficiently determine $\lambda_c$ throughout the phase diagram using a
single lattice volume at each point, and looking for the point in the
spectrum where $I_{s_0}$ takes the value $I_{s_0,c}$. We use again
natural cubic splines to interpolate the numerical data, using the
more precise determination of $I_{s_0,c}$ obtained in the deconfined
phase and generating synthetic data as discussed above to estimate the
statistical error. To estimate the magnitude of finite size effects,
we determined also the crossing points $\lambda_{c,\pm}$ of $I_{s_0}$
with $I_{s_0,c}\pm \delta I_{s_0,c}$, with $\delta I_{s_0,c}$ the
uncertainty on $I_{s_0,c}$.  This is meant to determine just how much
the crossing point of $I_{s_0}$ may change with the volume, as the
error band on $I_{s_0,c}$ is determined by the fluctuations of the
crossing point of the various pairs of volumes used to find the
mobility edge and the critical statistics in section
\ref{sec:mobedge}, and has nothing to do with the fact that
$I_{s_0,c}$ is not known exactly. As explained in section
\ref{sec:stloc}, one could in fact use any value intermediate between
the RMT and the Poisson expectations to give an estimate of the
mobility edge in a finite volume, and this would converge to the
correct value in the thermodynamic limit.

\begin{figure}[thb]
  \centering
  \includegraphics[width=0.45\textwidth]{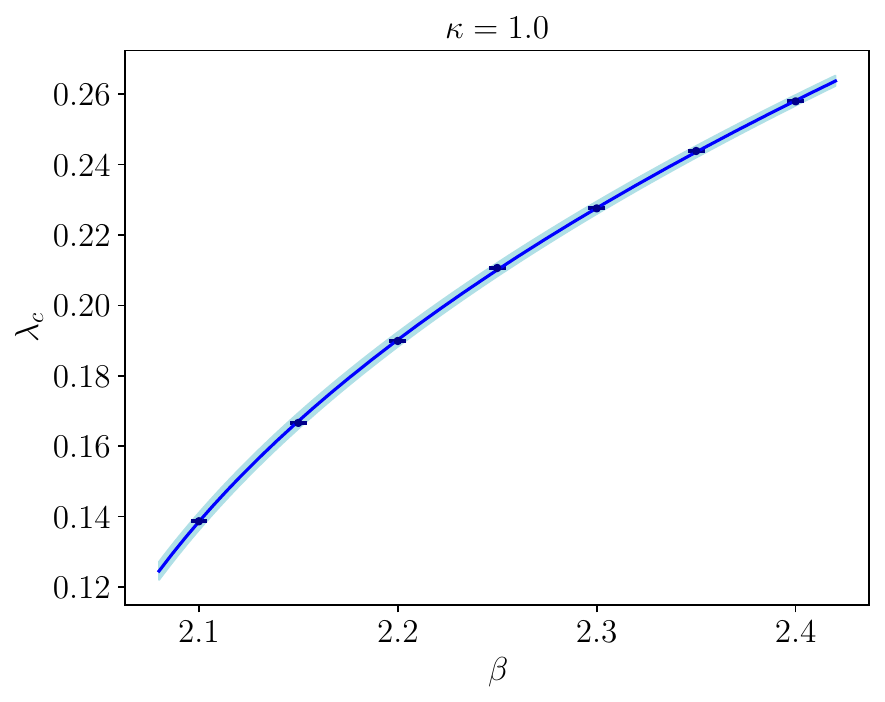}
  \caption{The dependence of the mobility edge on $\beta$ in the Higgs
    phase on the line at constant $\kappa=1.0$. The solid line is a
    power-law fit, Eq.~\eqref{eq:plaw}, to the numerical data; the
    band corresponds to the finite-size systematic uncertainty
    discussed in the text. The point where the mobility edge vanishes
    is estimated at $\beta_{\mathrm{loc}}=2.0101(28)$, in the
    crossover region between the confined and Higgs phases, see
    Fig.~\ref{fig:2_0} (center).}
  \label{fig:31}
\end{figure}
\begin{table}[b]
  \centering
  \begin{tabular}{c|cc}
    &    deconfined &  Higgs\\ \hline\hline
    $\beta_{\mathrm{loc}}$ & 2.2997(22) & 2.0101(25)\\
    $ A$ & 0.3836(17) & 0.3851(14)\\
    $B$ & 0.3592(54) & 0.4344(59)\\ \hline
                           $\chi^2/\mathrm{dof}$& 1.48 & 1.64
  \end{tabular}
  \caption{Parameters of a best fit of the $\beta$ dependence of the
    mobility edge in the deconfined ($\kappa=0.3$) and Higgs
    ($\kappa=1.0$) phases, with the fitting function in
    Eq.~\eqref{eq:plaw}. Only statistical errors are reported.}
  \label{tab:bc}
\end{table}
  
\begin{figure}[thb]
  \centering
  \includegraphics[width=0.45\textwidth]{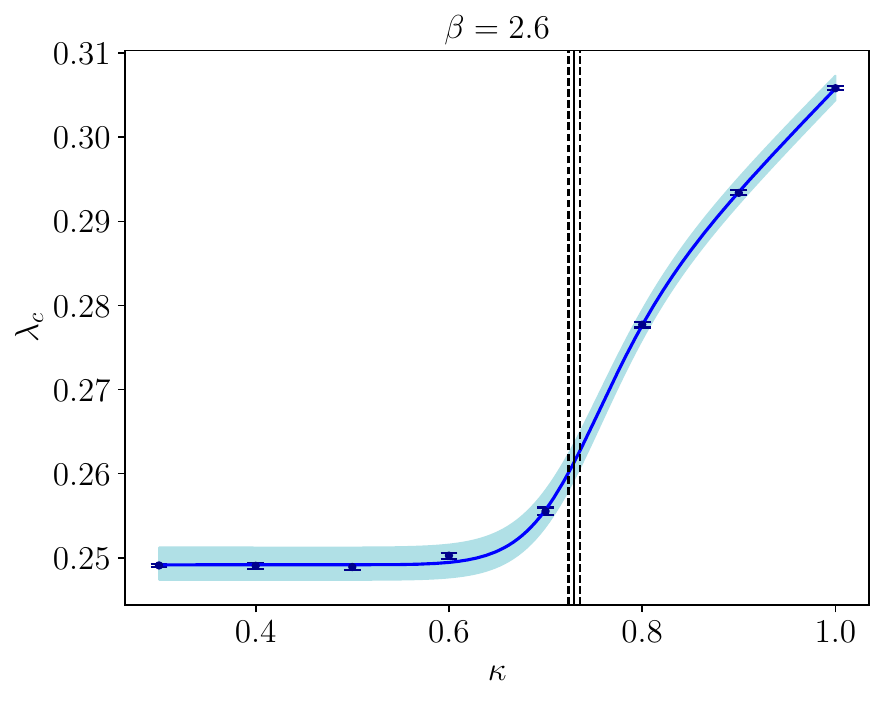}
  \caption{The dependence of the mobility edge on $\kappa$ on the line
    at constant $\beta=2.6$ in the deconfined and Higgs phases.  The
    solid line is a fit to the data with Eq.~\eqref{eq:sigmo}; the
    band corresponds to the finite-size systematic uncertainty
    discussed in the text. A change of behavior is found at
    $\kappa_{\mathrm{loc}}=0.7303(59)$ in the crossover region between
    the two phases, see Fig.~\ref{fig:2_0} (top), marked here by a
    vertical solid line, with dashed lines giving the corresponding
    error band.}
  \label{fig:32}
\end{figure}

We can then study how $\lambda_c$ depends on $\kappa$ and $\beta$.  In
Fig.~\ref{fig:30} we show how $\lambda_c$ changes in the deconfined
phase as one decreases $\beta$ towards the confined phase at fixed
$\kappa$. We expect that the mobility edge disappears as we enter the
confined phase and the Polyakov loop loses its strong ordering. To
estimate the value $\beta_{\mathrm{loc}}(\kappa)$ of $\beta$ where
this happens we fitted our results with a power-law function,
\begin{equation}
  \label{eq:plaw}
  \lambda_c(\beta)= A\cdot (\beta-\beta_{\mathrm{loc}})^{B}\,,
\end{equation}
using the MINUIT library~\cite{James:1975dr} to minimize the $\chi^2$,
computed using only the statistical errors on $\lambda_c$. We then
repeated the fit using $\lambda_{c\pm}(\beta)$ to find the
corresponding $\beta_{\mathrm{loc}\pm}$ where they extrapolate to
zero, and used these to estimate the systematic uncertainty due to
finite-size effects as
$\f{1}{2}|\beta_{\mathrm{loc}+}-\beta_{\mathrm{loc}-}|$. We obtained
for the critical value
$\beta_{\mathrm{loc}}(0.3) =
2.2997(22)_{\mathrm{stat}}(53)_{\mathrm{syst}}= 2.2997(57)$, where the
total error is the sum in quadrature of the statistical error from the
fit and of the systematic error.  The other fit parameters and the
$\chi^2$ per degree of freedom,
$\chi^2/\mathrm{dof}=
\chi^2/(n_{\mathrm{data}}-n_{\mathrm{parameters}})$, are reported in
Tab.~\ref{tab:bc}.  The critical point is shown also in
Fig.~\ref{fig:2_0} (bottom), where we see that the vanishing of the
mobility edge matches well with the crossover between the phases.

In Fig.~\ref{fig:31} we show how $\lambda_c$ changes in the Higgs
phase as one decreases $\beta$ towards the confined phase at fixed
$\kappa$. Again, we expect the mobility edge to disappear at the
crossover. For the critical $\beta_{\mathrm{loc}}(\kappa)$ we find
$\beta_{\mathrm{loc}}(1.0)=2.0101(25)_{\mathrm{stat}}(13)_{\mathrm{syst}}
=2.0101(28)$, again from a fit with a power law, Eq.~\eqref{eq:plaw},
using statistical errors only (see Tab.~\ref{tab:bc} for the other fit
parameters), and estimating systematic effects by fitting
$\lambda_{c\pm}$, as discussed above. This is shown also in
Fig.~\ref{fig:2_0} (center), where one sees that the vanishing of
$\lambda_c$ takes place again in the crossover region.

Instead of extrapolating in $\beta$, one could in principle explore
the crossover region directly without particular problems, as there is
no critical slowing down taking place there. However, our
extrapolations convincingly show that $\lambda_c$ will be close to
zero near the crossover to the confined phase. In the near-zero region
the spectral density is low, and the effects of the approximate taste
symmetry of staggered fermions on the spectrum is prominent. These
effects consist in the formation of nearly degenerate multiplets of
eigenvalues that distort the spectral statistics away from the
expected universal behavior (see Ref.~\cite{Kovacs:2012zq}), making
our method unreliable. To cure this problem one needs to make the
lattice volume large enough, so that the size of the would-be
multiplets (which is controlled by the lattice spacing) becomes larger
than the typical level spacing (which is controlled by the inverse
lattice volume), and the approximate taste symmetry does not affect
the short-range spectral statistics. This is numerically expensive,
and we have preferred to adopt here a computationally less intensive
method, leaving the direct investigation of the crossover region to
future work.

\begin{table}[b]
  \centering
    \begin{tabular}{c|c}
      \hline\hline
      $\kappa_{\mathrm{loc}}$ & 0.7303(56)\\
      $a$ & 0.24915(13)\\
      $b$ & 0.1185(53)\\
      $c$ & 0.1874(52)\\
      $d$ & 26.4(2.0) \\
      \hline
      $\chi^2/\mathrm{dof}$& 2.05
  \end{tabular}
  \caption{Parameters of a best fit of the $\kappa$ dependence of the
    mobility edge in the deconfined and Higgs phases at $\beta=2.6$,
    with the fitting function in Eq.~\eqref{eq:sigmo}. Only
    statistical errors are reported.}
  \label{tab:kc}
\end{table}

The third case we examined is the transition from the Higgs phase to
the deconfined phase as we decrease $\kappa$ at fixed $\beta$. This is
shown in Fig.~\ref{fig:32}. One can see that at first $\lambda_c$
decreases quickly with $\kappa$, but below a critical value
$\kappa_{\mathrm{loc}}(\beta)$ it becomes practically constant. The
critical value is defined here as the point where the behavior changes
from approximately constant to approximately linear, as obtained by
fitting with the following function,
\begin{equation}
  \label{eq:sigmo}
  \begin{aligned}
  \lambda_c (\kappa) &=
  a\cdot\left(1-\sigma\left(d\cdot(\kappa-\kappa_{\mathrm{loc}})\right)\right)
  \\ & \phantom{=}+
  \left(b\kappa+c\right)\sigma(d\cdot(\kappa-\kappa_{\mathrm{loc}}))\,,  
  \end{aligned}
\end{equation}
where $\sigma(x)={1}/{(1+e^{-x})}$ is the sigmoid function.  Following
the same procedure discussed above to estimate errors, we found
$\kappa_{\mathrm{loc}}(2.6)=
0.7303(57)_{\mathrm{stat}}(17)_{\mathrm{syst}} = 0.7303(59)$ for the
critical point (see Tab.~\ref{tab:kc} for the other fit
parameters). As shown in Fig.~\ref{fig:2_0} (top), also in this case
the critical value matches well with the position of the
crossover. Notice that here the critical point is not as sharply
defined as in the previous two cases, as it simply corresponds to a
change in the $\kappa$ dependence of the mobility edge, rather than
its very appearance. However, it is possible that the change in the
behavior of $\lambda_c(\kappa)$ becomes singular in the
infinite-volume limit, e.g., due to a discontinuity in its
derivative. If so, one would find a sharply defined critical point for
the geometric transition also in this case. At the present stage this
is only speculation, and a more careful determination of the mobility
edge is needed to test this possibility, either by a proper
finite-size scaling analysis, or by checking the volume dependence of
the crossing point of $I_{s_0}$ with $I_{s_0,c}$.

\begin{figure}[h!]
  \centering

  \includegraphics[width=0.4\textwidth]{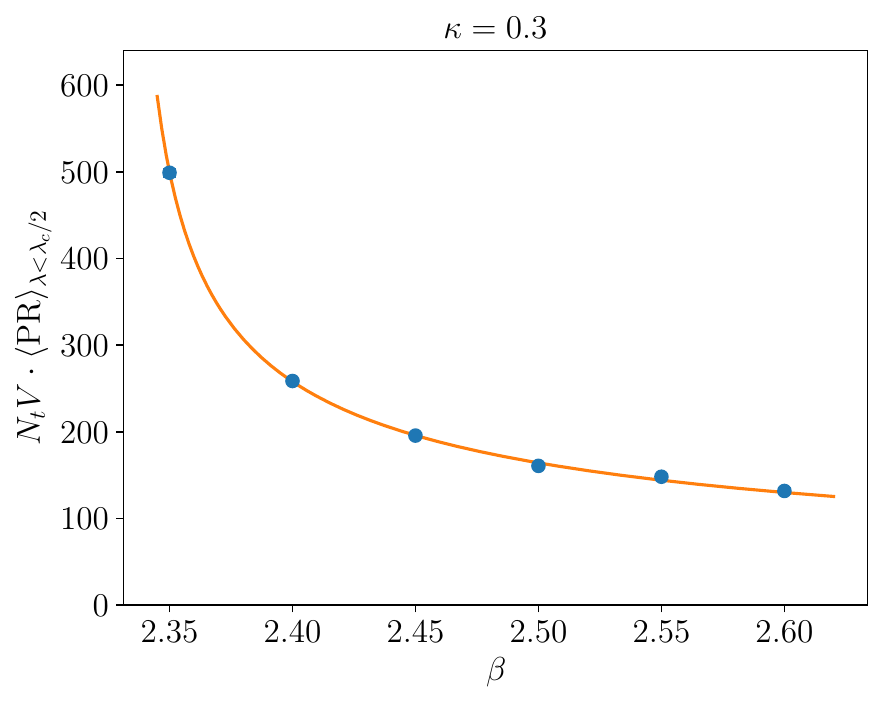}
  
  \includegraphics[width=0.4\textwidth]{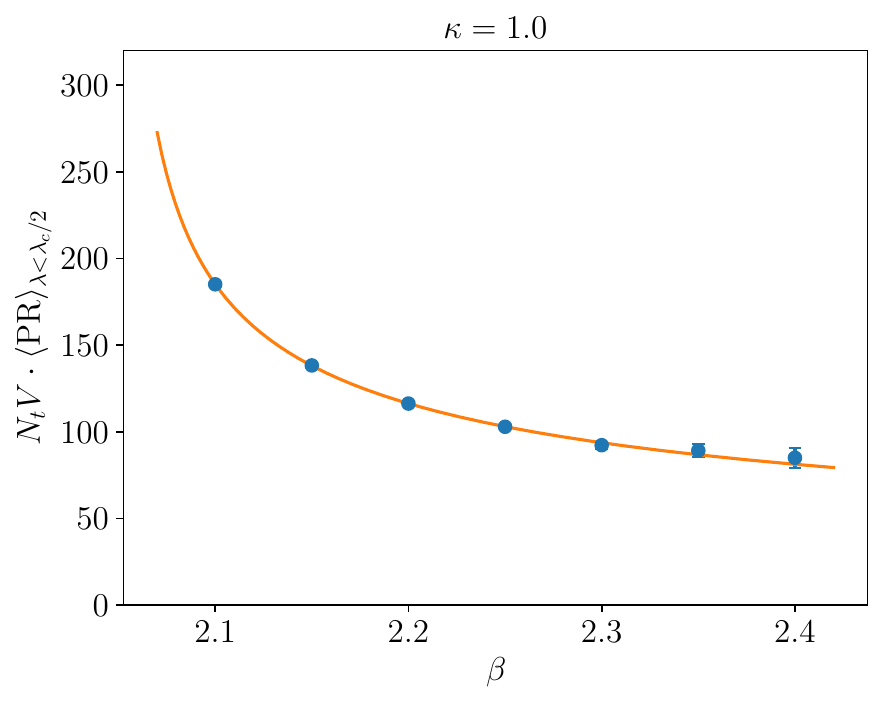}

    \includegraphics[width=0.4\textwidth]{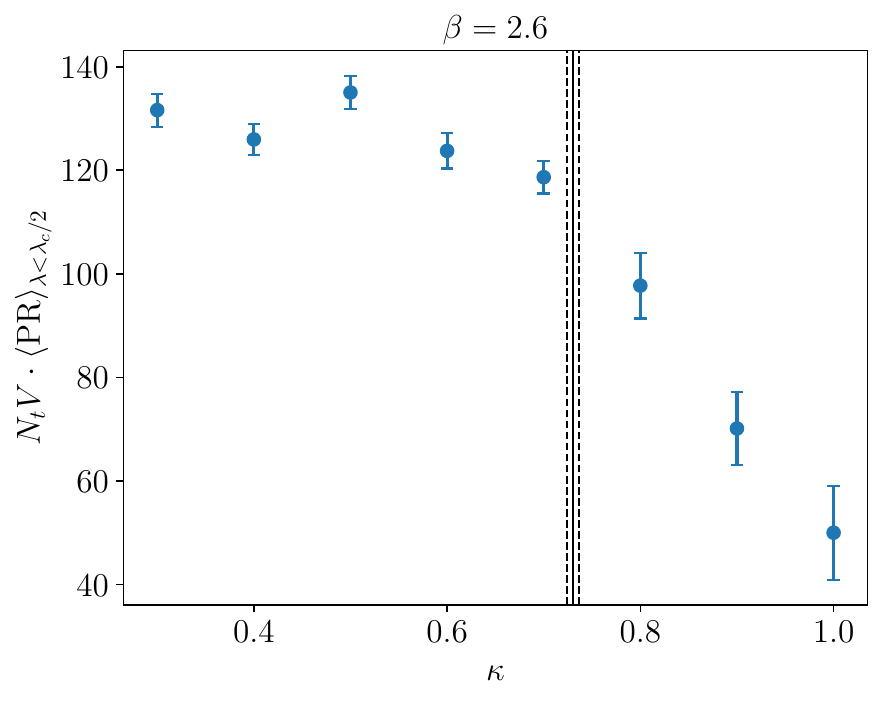}
    \caption{Mode size averaged up to $\lambda_c/2$,
      Eq.~\eqref{eq:avPRdef}, at $\kappa=0.3$ in the deconfined phase
      (top panel) and at $\kappa=1.0$ in the Higgs phase (center
      panel), as a function of $\beta$, and at $\beta=2.6$ across the
      transition from the deconfined to the Higgs phase (bottom
      panel), as a function of $\kappa$. In all plots $N_s=20$ and
      $N_t=4$. The solid line in the top and center panels is a fit
      with a power-law function. The vertical and dashed lines in the
      bottom panel mark the critical value $\kappa_{\mathrm{loc}}$ and
      the corresponding error band [see after Eq.~\eqref{eq:sigmo}].}
  \label{fig:avPR}
\end{figure}

\begin{table}[b]
  \centering
  \begin{tabular}{c|cc}
    & deconfined & Higgs\\ \hline
    $\beta_{\mathrm{loc}}$ & $2.3318(15)$    & $2.0499(92) $\\
    $a$ & $67.2(2.1)$       & $52.0(2.9)$\\
    $b$ & $0.501(17)$      & $0.424(43)$\\
    \hline
    $\chi^2/\mathrm{dof}$ & 1.96 & 0.34

  \end{tabular}
  \caption{Parameters of a best fit of the $\beta$ dependence of the
    average size of the lowest modes,
    $\la N_t V \cdot \mathrm{PR}\ra_{\lambda<\f{\lambda_c}{2}}$,
    Eq.~\eqref{eq:avPRdef}, with the fitting function in
    Eq.~\eqref{eq:avPRfit}.}
  \label{tab:aa}
\end{table}

It is interesting to compare the estimates of $\beta_{\mathrm{loc}}$
and $\kappa_{\mathrm{loc}}$ obtained from eigenvalue observables to
similar estimates obtained from eigenvector observables. In
particular, if $\lambda_c$ vanishes continuously at
$\beta_{\mathrm{loc}}$, then in the thermodynamic limit the
localization length of the low modes should correspondingly
diverge. We have then looked at the size of the low modes averaged
over the lowest half of the localized spectral region,
\begin{equation}
  \label{eq:avPRdef}
  \begin{aligned}
  N_t V \cdot  \la \mathrm{PR} \ra_{\lambda<\f{\lambda_c}{2}} &=
  \f{1}{\mathcal{N}(\tf{\lambda_c}{2})}\int_0^{\f{\lambda_c}{2}} d\lambda
  \,\rho(\lambda)\, N_t V \cdot\mathrm{PR}(\lambda,N_s) \,,\\
  \mathcal{N}(\lambda_0)&= \int_0^{\lambda_0}
  d\lambda \,\rho(\lambda)\,.
  \end{aligned}
\end{equation}
In Fig.~\ref{fig:avPR} we show this quantity as a function of $\beta$
for constant $\kappa=0.3$ in the deconfined phase (top panel) and
$\kappa=1.0$ in the Higgs phase (center panel). This quantity does
indeed grow large as one approaches the confined phase. Fits with a
power-law function,
\begin{equation}
  \label{eq:avPRfit}
  N_t V \cdot  \la \mathrm{PR}\ra_{\lambda<\f{\lambda_c}{2}}= a\cdot
  (\beta-\beta_{\mathrm{loc}})^{-b} \,, 
\end{equation}
yield $\beta_{\mathrm{loc}}=2.3318(15)$ in the deconfined phase, and
$\beta_{\mathrm{loc}}=2.0499(92)$ in the Higgs phase, both in the
crossover region, and in reasonable agreement with the determinations
based on the extrapolation of the mobility edge. Here one should take
into account that the functional form Eq.~\eqref{eq:avPRfit} is not
fully justified, as the mode size cannot diverge in a finite volume,
and there is no reason to assume that the mode size goes to zero at
large $\beta$. Nonetheless, one obtains decent fits (see the resulting
fit parameters and $\chi^2$ in Tab.~\ref{tab:aa}); adding a constant
term makes them worse. On top of this, the error estimates do not
include any uncertainty due to finite-size effects, which are large
near $\beta_{\mathrm{loc}}$. For completeness, in the bottom panel of
Fig.~\ref{fig:avPR} we show
$ N_t V \cdot \la \mathrm{PR} \ra_{\lambda<\f{\lambda_c}{2}}$ as a
function of $\kappa$ at constant $\beta=2.6$ across the two
phases. Here the data indicate a finite mode size at all $\kappa$,
with a change from a constant to a steadily decreasing trend taking
place at the crossover between the deconfined and the Higgs phase,
showing that localized modes shrink rapidly as one moves deeper in the
Higgs phase and the Polyakov-loop expectation value increases (see
Fig.~\ref{fig:0}).

\subsection{Correlation with bosonic observables and sea/island mechanism}
\label{sec:bos}

\begin{figure}[t!]
  \centering
  \includegraphics[width=0.4\textwidth]{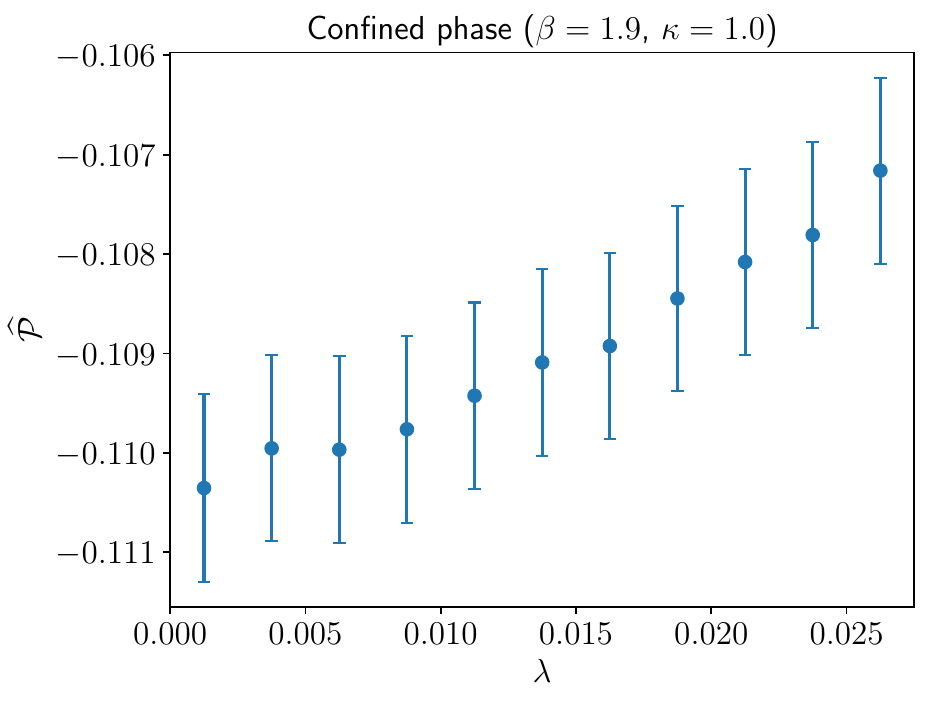}

  \includegraphics[width=0.4\textwidth]{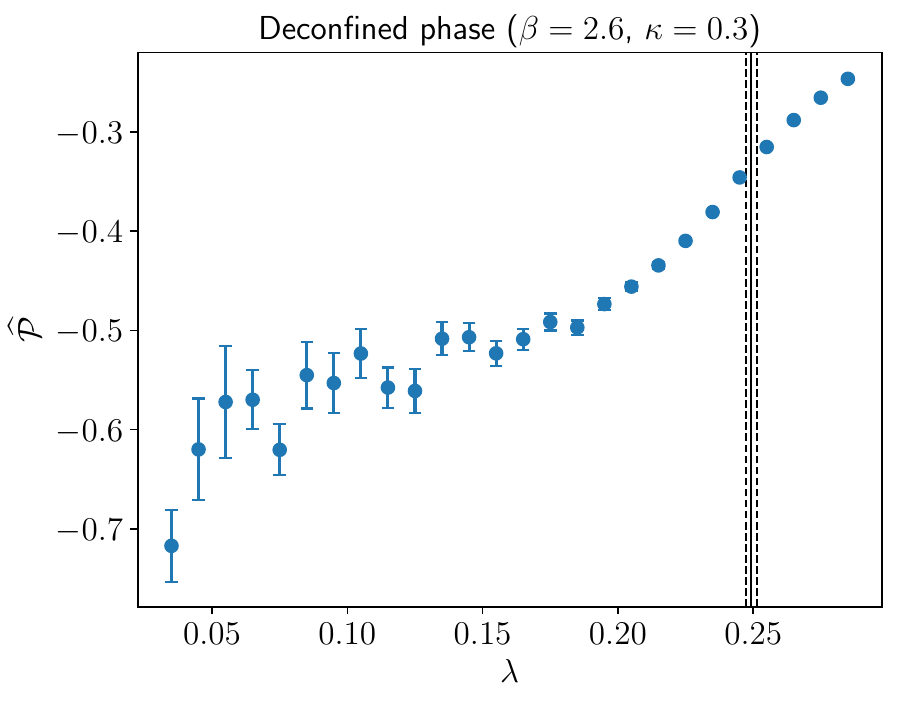}
  
  \includegraphics[width=0.4\textwidth]{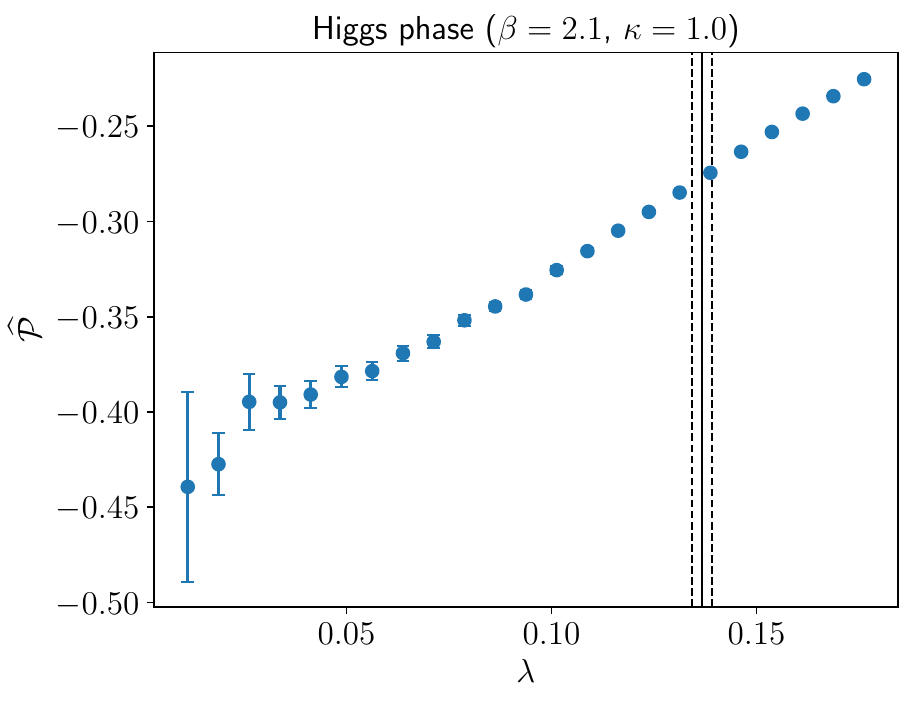}
  \caption{Polyakov loop weighted by Dirac modes, centered to its
    average and rescaled by the square root of its susceptibility,
    Eq.~\eqref{eq:su2h15bis}, at $\beta=1.9$ and $\kappa=1.0$ in the
    confined phase (top panel; here $N_s=20$), at $\beta=2.6$ and
    $\kappa=0.3$ in the deconfined phase (center panel; here
    $N_s=32$), and at $\beta=2.1$ and $\kappa=1.0$ in the Higgs phase
    (bottom panel; here $N_s=32$). In all plots $N_t=4$. In the center
    and bottom panels the solid line shows the value of the mobility
    edge, and the dashed lines indicate the corresponding error band.}
  \label{fig:new1}
\end{figure}
\begin{figure}[t!]
  \centering
  \includegraphics[width=0.4\textwidth]{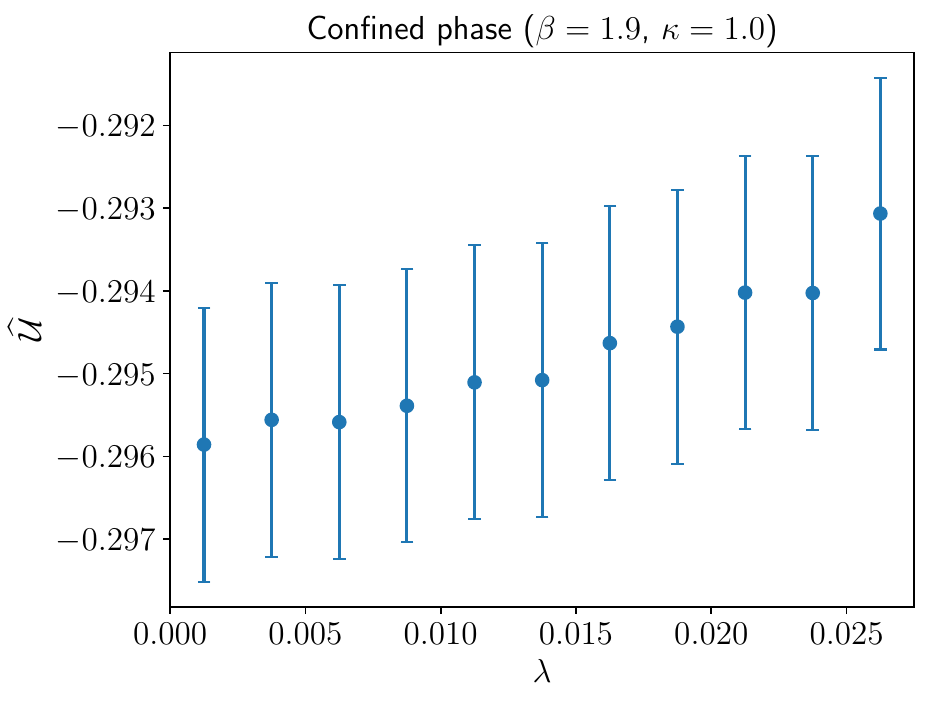}
  
  \includegraphics[width=0.4\textwidth]{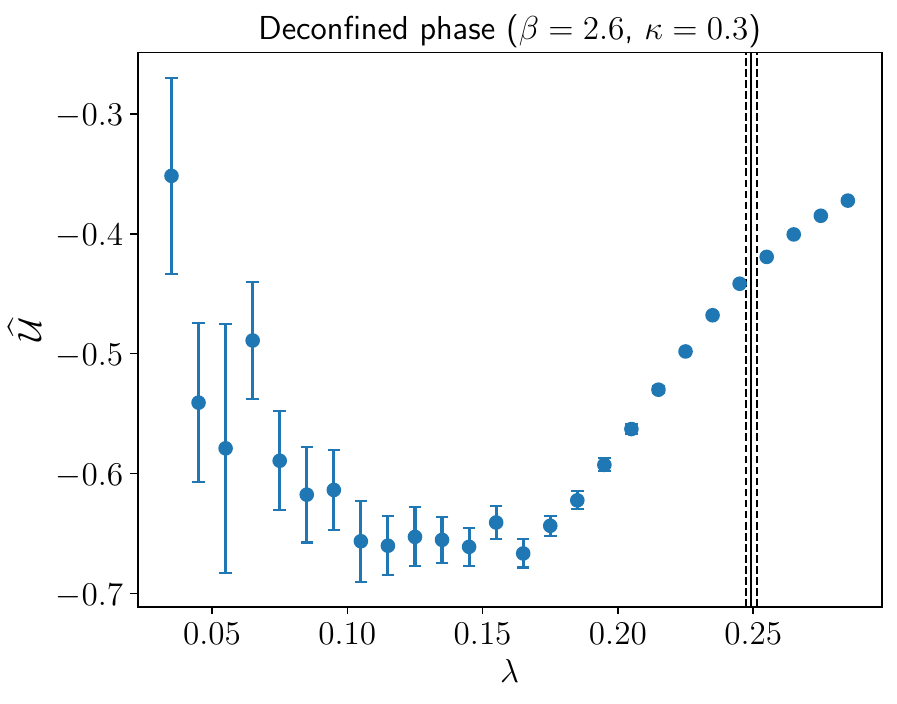}\hfil
  \includegraphics[width=0.4\textwidth]{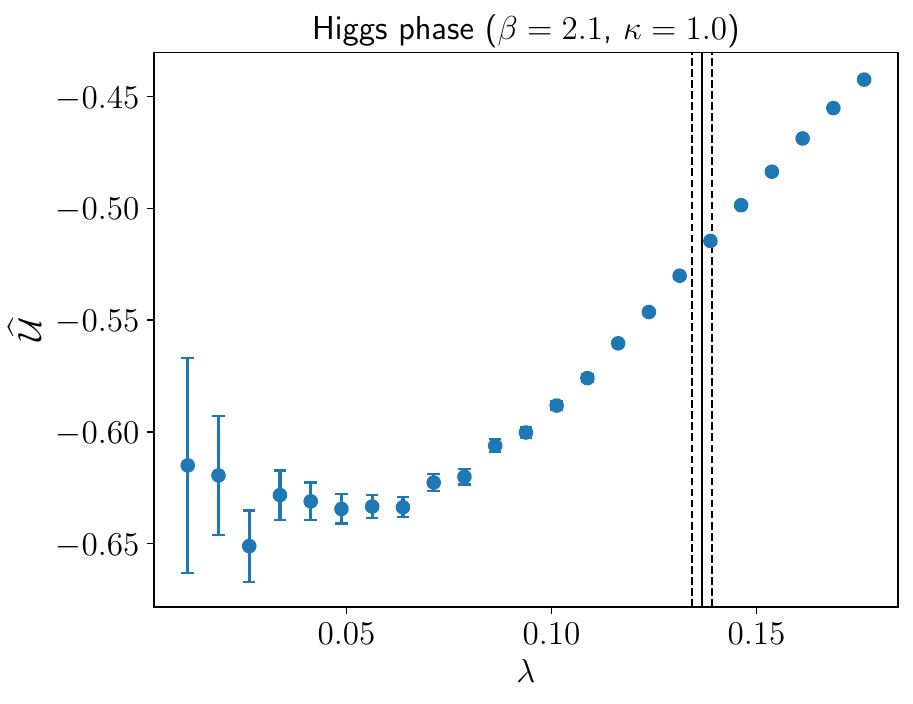}
  \caption{Plaquette weighted by Dirac modes, centered to its average
    and rescaled by the square root of its susceptibility,
    Eq.~\eqref{eq:su2h15bis}, at $\beta=1.9$ and $\kappa=1.0$ in the
    confined phase (top panel; here $N_s=20$), at $\beta=2.6$ and
    $\kappa=0.3$ in the deconfined phase (center panel; here
    $N_s=32$), and at $\beta=2.1$ and $\kappa=1.0$ in the Higgs phase
    (bottom panel; here $N_s=32$). In all plots $N_t=4$. In the center
    and bottom panels the solid line shows the value of the mobility
    edge, and the dashed lines indicate the corresponding error band.}
  \label{fig:new2}
\end{figure}
\begin{figure}[t!]
  \centering
  \includegraphics[width=0.4\textwidth]{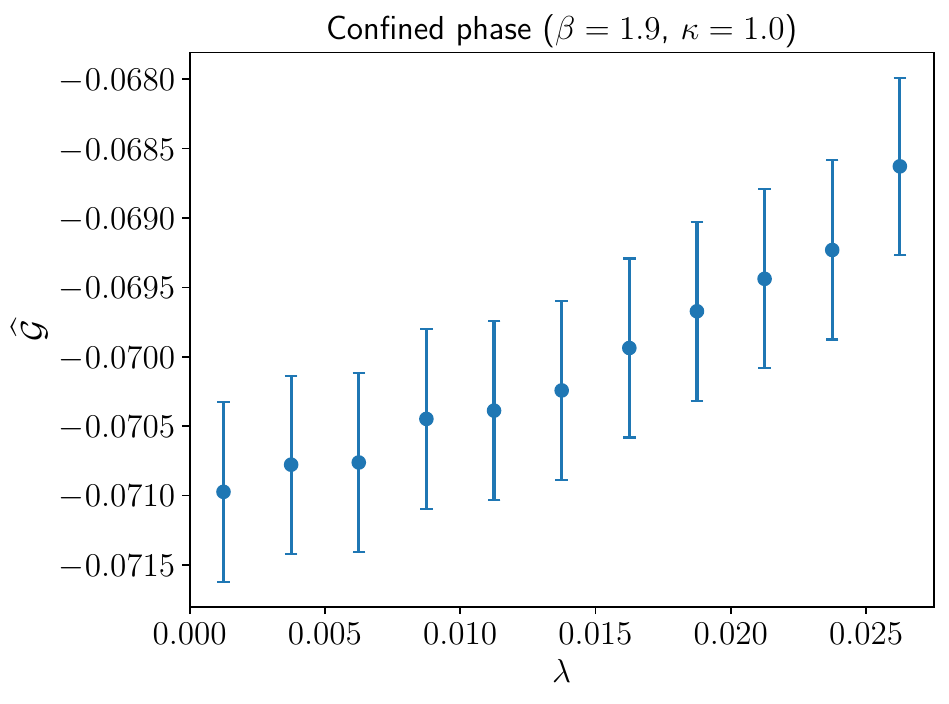}
  
  \includegraphics[width=0.4\textwidth]{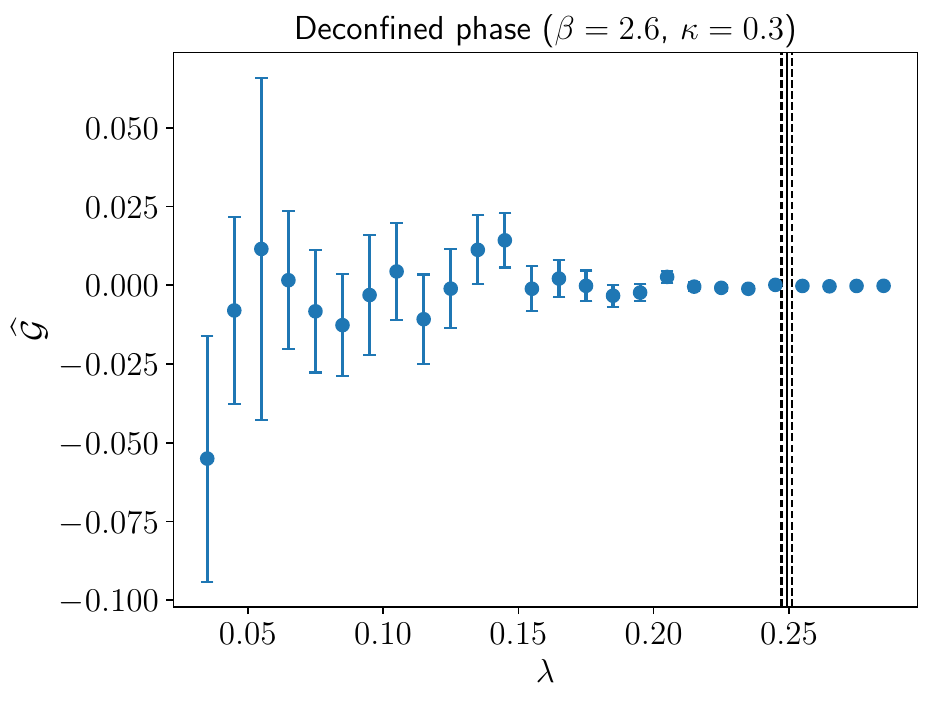}\hfil
  \includegraphics[width=0.4\textwidth]{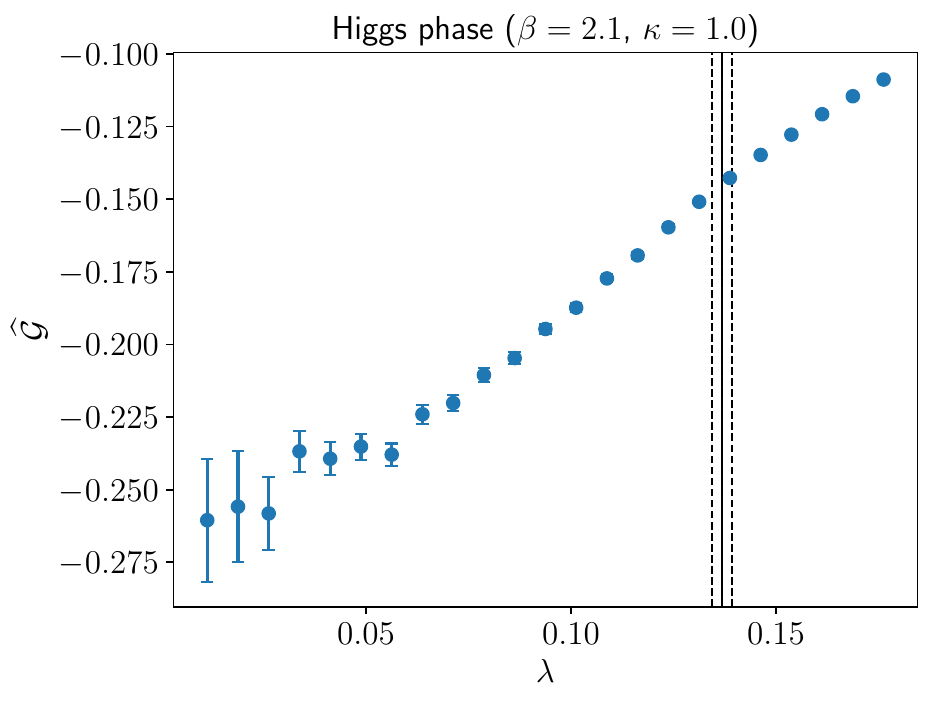}
  \caption{Gauge-Higgs coupling term weighted by Dirac modes, centered
    to its average and rescaled by the square root of its
    susceptibility, Eq.~\eqref{eq:su2h15bis}, at $\beta=1.9$ and
    $\kappa=1.0$ in the confined phase (top panel; here $N_s=20$), at
    $\beta=2.6$ and $\kappa=0.3$ in the deconfined phase (center
    panel; here $N_s=32$), and at $\beta=2.1$ and $\kappa=1.0$ in the
    Higgs phase (bottom panel; here $N_s=32$). In all plots
    $N_t=4$. In the center and bottom panels the solid line shows the
    value of the mobility edge, and the dashed lines indicate the
    corresponding error band.}
  \label{fig:new3}
\end{figure}

We now proceed to discuss our results on the correlation of staggered
eigenmodes with the gauge and Higgs fields. To this end, the most
informative quantities are the centered and normalized observables
$\widehat{\mathcal{U}}$, $\widehat{\mathcal{P}}$, and
$ \widehat{\mathcal{G}}$, defined in Eq.~\eqref{eq:su2h15bis}, that
take into account the width of the distribution of the relevant
bosonic observables. Our results for these quantities are shown in
Figs.~\ref{fig:new1}--\ref{fig:new3}. The statistical error on the
numerical estimate of these quantities is obtained by first
determining the jackknife error on $\mathcal{U}$, $\mathcal{P}$, and
$\mathcal{G}$, and correspondingly on $\la U\ra$, $\la P\ra$,
$\la G\ra$ and on $(\delta U)^2$, $(\delta P)^2$, $(\delta G)^2$,
followed by linear error propagation. Correlations with Polyakov-loop
and plaquette fluctuations are always negative, showing that low modes
prefer locations where these quantities fluctuate to values below
their average. Correlations with gauge-Higgs coupling term
fluctuations are again negative in the confined and in the Higgs
phase, while they are essentially compatible with zero in the
deconfined phase.

The correlation of low modes with Polyakov-loop fluctuations is shown
in Fig.~\ref{fig:new1}. In the confined phase this is small but
significant, and decreasing very little in magnitude as one goes up in
the spectral region that we explored. The strength of this correlation
is considerably larger in the Higgs phase, and even larger in the
deconfined phase. Since Polyakov-loop fluctuations are typically
localized in these phases, this increased correlation is possible only
if the low modes tend to localize on the corresponding locations. In
both the deconfined and the Higgs phase one sees also a more rapid
decrease in the magnitude of the correlation as one moves up in the
spectrum. This, however, remains stronger than for the lowest modes in
the confined phase also above the mobility edge.

The correlation of low modes with plaquette fluctuations is shown in
Fig.~\ref{fig:new2}. Also in this case a significant correlation is
found in all three phases, generally stronger (and comparable in size)
in the deconfined and Higgs phases than in the confined
phase. Compared to the correlation with Polyakov-loop fluctuations,
one finds a similar magnitude in the deconfined phase, and a larger
magnitude in the Higgs phase. Since also plaquette fluctuations are
typically localized, this means that they are at least as relevant as
Polyakov-loop fluctuations for the localization of low modes.  A clear
upturn is visible for the lowest modes in the deconfined phase and, to
a much smaller extent, also in the Higgs phase.  We do not have an
explanation for this.  Even though the density of near-zero modes is
very small in both cases, leading to large fluctuations, this upturn
might be significant, as the mode size displays a similar behavior
(see Figs.~\ref{fig:20} and \ref{fig:22}), with an increase in size
for the lowest modes. (The downturn seen in $I_{s_0}$,
Figs.~\ref{fig:27} and \ref{fig:28}, may also be related, but could
also be a finite-size artifact caused by the low and rapidly changing
density of modes, that makes our unfolding procedure not fully
reliable in that spectral region.) The same upturn in the mode size is
observed also in QCD~\cite{Kovacs:2012zq}, where it can be explained
by the topological origin of the near-zero
modes~\cite{Edwards:1999zm,Vig:2021oyt}. Such modes are in fact
expected to originate in the mixing of the localized zero modes
associated with topological lumps in the gauge configuration at finite
temperature, so extending over more than one such lump. While they
fail to become delocalized due to the low density of lumps at high
temperature, they nonetheless should display a larger size than
localized modes not of topological origin. This picture is consistent
with the strong correlation between localized near-zero modes and the
local topology of the gauge configuration, demonstrated in
Ref.~\cite{Cossu:2016scb}, and with the lumpy nature of near-zero
Dirac modes in high-temperature QCD, demonstrated in
Ref.~\cite{Meng:2023nxf}. A similar mechanism could explain the larger
size of the lowest modes observed here. Interestingly, no upturn in
the size of the lowest modes is observed in 2+1 dimensional pure
$\mathrm{SU}(3)$ gauge theory~\cite{Giordano:2019pvc} or in 2+1
dimensional discrete gauge
theories~\cite{Baranka:2021san,Baranka:2022dib}, where the topology of
gauge field configurations is trivial.

Finally, the correlation of low modes with fluctuations of the
gauge-Higgs coupling term is shown in Fig.~\ref{fig:new3}. A very mild
correlation is visible in the confined phase, no significant
correlation is found in the deconfined phase, and a clear but small
correlation is found in the Higgs phase, weaker than the correlation
with Polyakov-loop and plaquette fluctuations. This leads us to
conclude that these fluctuations are much less relevant to low-mode
localization.

\begin{figure}[t!]
  \centering
  \includegraphics[width=0.37\textwidth]{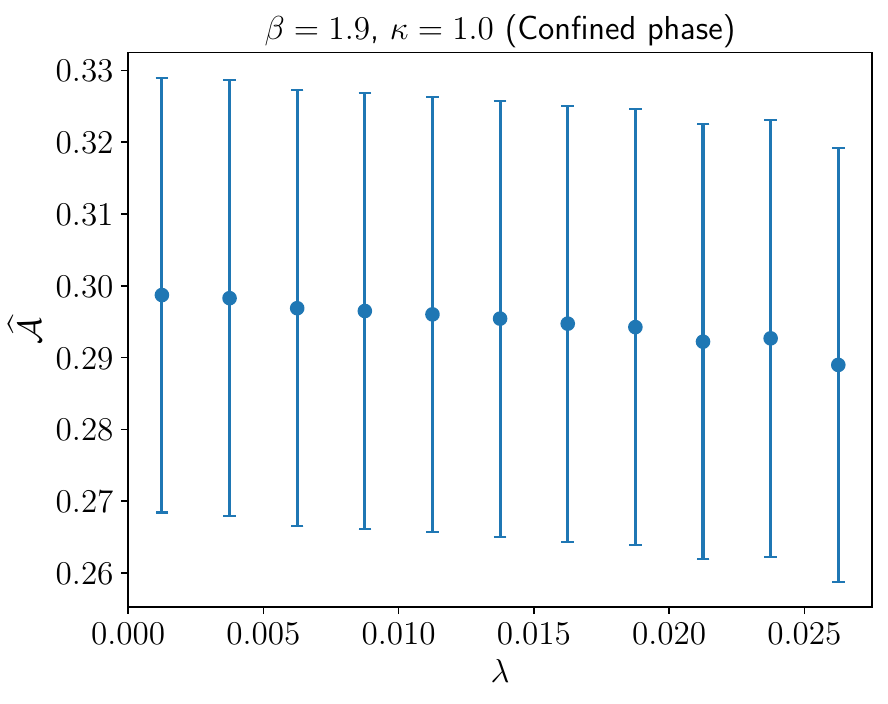}
  
  \includegraphics[width=0.38\textwidth]{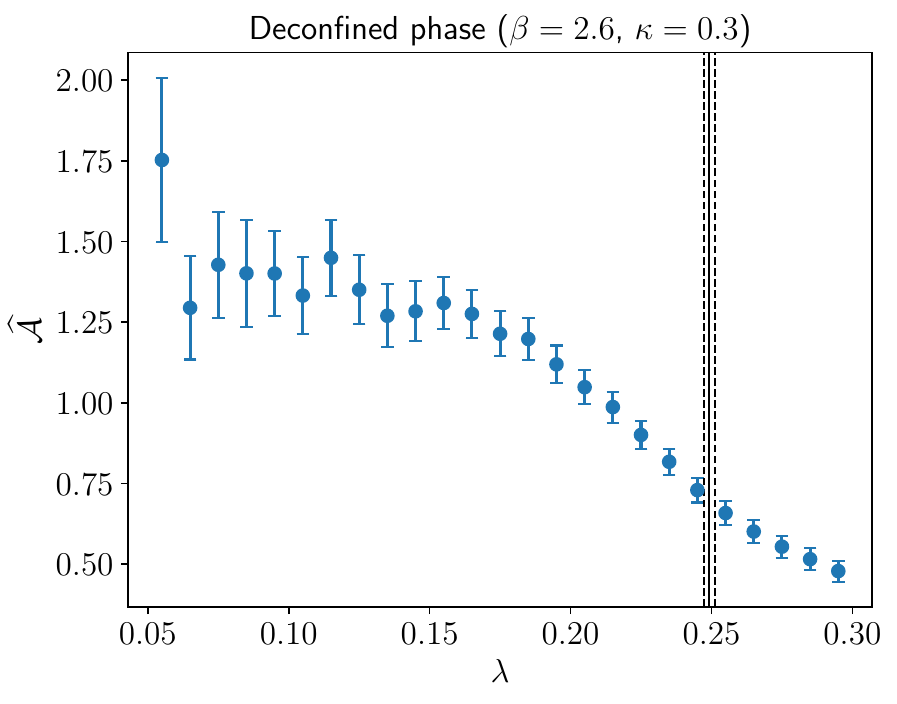}\hfil
  \includegraphics[width=0.38\textwidth]{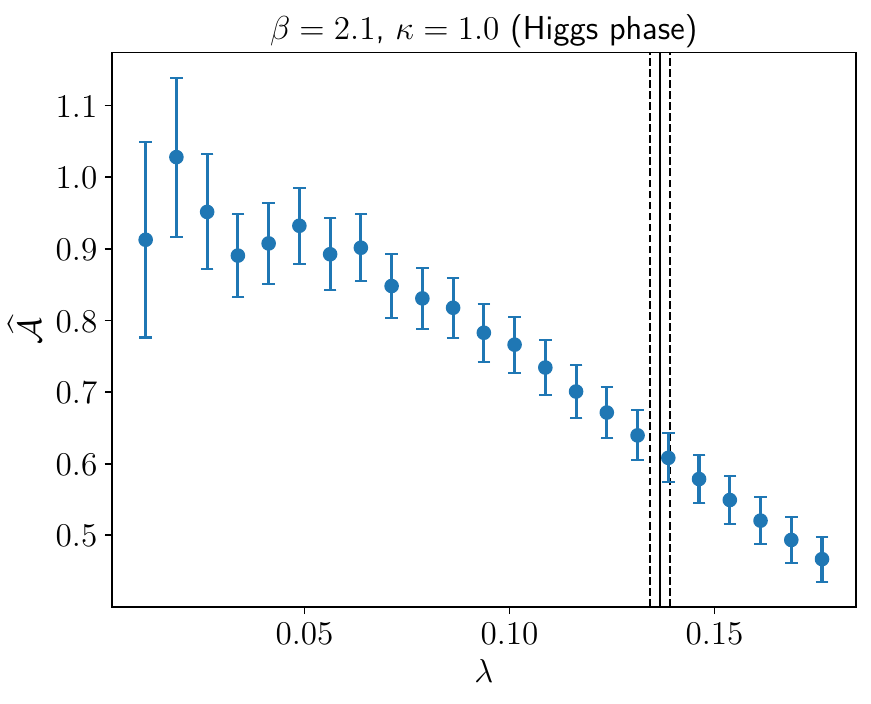}
  \caption{The quantity $\widehat{\mathcal{A}}$, Eq.~\eqref{eq:si8},
    measuring the correlation of staggered modes with fluctuations of
    $A(\vec{x})$, Eq.~\eqref{eq:si7}, at $\beta=1.9$ and $\kappa=1.0$
    in the confined phase (top panel; here $N_s=16$), at $\beta=2.6$
    and $\kappa=0.3$ in the deconfined phase (center panel; here
    $N_s=20$), and at $\beta=2.1$ and $\kappa=1.0$ in the Higgs phase
    (bottom panel; here $N_s=20$). In all plots $N_t=4$. In the center
    and bottom panels the solid line shows the value of the mobility
    edge, and the dashed lines indicate the corresponding error band.}
  \label{fig:62bis}
\end{figure}
  
We then studied the sea/island mechanism directly by looking at the
correlation of the staggered eigenmodes with the local fluctuations of
the hopping term in the Dirac-Anderson Hamiltonian, measured by the
quantity $A$ of Eq.~\eqref{eq:si6}. To this end we analyzed 450
configurations with $N_s=16$ in the confined phase, and 1400
configurations with $N_s=20$ in the deconfined and Higgs phases, with
$N_t=4$ in both cases.  The average value of $A$ drops substantially
as one moves from the confined to the deconfined or to the Higgs
phase: for the given lattice sizes (but this quantity is not expected
to show a strong volume dependence), $\la A\ra = 0.2761(11)$ at
$\beta=1.9,\kappa=1.0$ in the confined phase; $\la A\ra = 0.15828(64)$
at $\beta=2.6, \kappa=0.3$ in the deconfined phase; and
$\la A\ra = 0.20518(86)$ at $\beta=2.1, \kappa=1.0$ in the Higgs
phase. This is expected to happen, as a consequence of the ordering of
the Polyakov loop and the resulting strong correlation in the temporal
direction~\cite{Baranka:2022dib}.

The centered and normalized quantity $\widehat{\mathcal{A}}$ defined
in Eq.~\eqref{eq:si8} is shown in Fig.~\ref{fig:62bis}.  This quantity
correlates positively with the spatial density of low modes in all
phases, in agreement with the refined sea/islands picture of
Ref.~\cite{Baranka:2022dib}. In the confined phase the magnitude of
the correlation with fluctuations in this quantity is comparable with
the correlation with plaquette fluctuations, and independent of the
position in the spectrum in the available region, within errors. In
the Higgs and, especially, in the deconfined phase this correlation is
much stronger than those with Polyakov-loop and with plaquette
fluctuations. Although it remains strong also at the beginning of the
bulk region, it reduces by about a third when going from the lowest
modes to the first delocalized modes right above the mobility edge.
Since fluctuations of $A(\vec{x})$ are typically localized in the
deconfined and Higgs phases, this result strongly suggests that they
are the ones mainly responsible for trapping the eigenmodes in space.

\section{Conclusions}
\label{sec:concl}

A strong connection has emerged in recent years between the
deconfinement phase transition in gauge theories with or without
fermionic matter, and the change in the localization properties of low
Dirac modes~\cite{GarciaGarcia:2006gr,Kovacs:2012zq,Giordano:2013taa,
  Ujfalusi:2015nha,Cossu:2016scb,Holicki:2018sms,Kehr:2023wrs,
  Giordano:2021qav,Gockeler:2001hr,Gattringer:2001ia,Gavai:2008xe,
  Kovacs:2009zj,Kovacs:2010wx,Bruckmann:2011cc,Giordano:2015vla,
  Giordano:2016cjs, Giordano:2016vhx, Kovacs:2017uiz,
  Giordano:2016nuu,Giordano:2019pvc,Vig:2020pgq,Bonati:2020lal,
  Cardinali:2021fpu,Baranka:2021san,Baranka:2022dib}. In this paper we
extended this line of research by studying the lattice
$\mathrm{SU}(2)$ Higgs model with a Higgs field of fixed
length~\cite{Fradkin:1978dv,Lang:1981qg,Montvay:1984wy,
  Montvay:1985nk,Langguth:1985eu,Campos:1997dc,Bonati:2009pf} at
finite temperature, probed with external static fermions. The
extension is twofold. On the one hand, this model has dynamical scalar
rather than fermionic matter: while one still expects localized modes
in the deconfined phase of the model, as the nature of the dynamical
matter does not affect the general argument for
localization~\cite{Bruckmann:2011cc,
  Giordano:2015vla,Giordano:2016cjs,Giordano:2016vhx,Giordano:2021qav,
  Baranka:2022dib}, it is nonetheless useful to verify this
explicitly. On the other hand, and more interestingly, the
two-parameter phase diagram of this model displays a third phase
besides the confined and deconfined phases, i.e., the Higgs phase: one
can then check whether or not modes are localized in this phase, and
if so whether the onset of localization is related in any way to the
thermodynamic transition.

A survey of the phase diagram shows the expected tripartition into a
confined, a deconfined, and a Higgs phase, separated by analytic
crossovers~\cite{Bonati:2009pf}. The deconfined and the Higgs phases
are distinguished from the confined phase by a much larger expectation
value of the Polyakov loop, and from each other by the expectation
value of the Higgs-coupling term, much larger in the Higgs phase than
in the deconfined and in the confined phases. Since the Polyakov loop
is strongly ordered, one expects localization of low Dirac modes to
take place in both phases~\cite{Bruckmann:2011cc,
  Giordano:2015vla,Giordano:2016cjs,Giordano:2016vhx,Giordano:2021qav,
  Baranka:2022dib}.

By means of numerical simulations, we have demonstrated that localized
modes are indeed present both in the deconfined and in the Higgs
phase. In both cases, the mobility edge separating localized and
delocalized modes in the spectrum decreases as one moves towards the
confined phase, and extrapolates to zero as one reaches the crossover
region. At the transition between the deconfined and the Higgs phase,
instead, the dependence of the mobility edge on the gauge-Higgs
coupling constant changes from almost constant to steadily
increasing. These findings provide further support to the universal
nature of the sea/islands picture of
localization~\cite{Bruckmann:2011cc,Giordano:2015vla,Giordano:2016cjs,
  Giordano:2016vhx,Giordano:2021qav, Baranka:2022dib} in a previously
unexplored setup in the presence of dynamical scalar matter.

We have then studied the sea/islands mechanism in more detail,
measuring the correlation between localized modes and fluctuations of
the gauge and Higgs fields. We found a strong correlation with
Polyakov-loop and plaquette fluctuations both in the deconfined and in
the Higgs phase, and a mild but significant correlation with
fluctuations of the gauge-Higgs coupling term only in the Higgs
phase. Moreover, we found in both phases a very strong correlation
(stronger than that with Polyakov-loop or plaquette fluctuations) with
the type of gauge-field fluctuations identified in
Ref.~\cite{Baranka:2022dib} as the most relevant to localization. This
provides further evidence for the validity of the refined sea/islands
picture proposed in Ref.~\cite{Baranka:2022dib}.

A straightforward extension of this work would be the direct study of
the region near the crossover to the confined phase, where larger
volumes are required to apply our method than the ones employed here,
in order to avoid the distortion effects of the approximate taste
symmetry of staggered fermions on the spectral
statistics~\cite{Kovacs:2012zq}. Another possible extension would be a
study of the low $\beta$, large $\kappa$ corner of the phase diagram,
where the crossover becomes very weak, in order to check if the line
of ``geometric'' transitions where the mobility edge in the Dirac
spectrum vanishes extends all the way to $\beta=0$, or if instead it
has an endpoint.  This is interesting also in connection with the
``spin glass'' approach of Ref.~\cite{Greensite:2021fyi}: since in
that region of parameter space this predicts a transition line clearly
distinct from the one found with more traditional approaches based on
gauge fixing, one would like to compare this line with the one defined
by the vanishing of the mobility edge (if the latter exists).  A
different direction would be the study of the localization properties
of the eigenmodes of the covariant Laplacian, extending to finite
temperature and dynamical scalar matter the work of
Refs.~\cite{Greensite:2005yu,Greensite:2006ns}.

\protect{\mbox{}}

\section*{Acknowledgments}
We thank T.~G.~Kov\'acs for useful discussions and a careful reading of
the manuscript, and I.~Varga for useful discussions. MG was partially
supported by the NKFIH grant KKP-126769. This work was also supported
by the NKFIH excellence Grant TKP2021-NKTA-64.

\bibliographystyle{apsrev4-2}
\bibliography{references_su2H}

\end{document}